\documentclass[printer]{aa} 

\usepackage{subcaption}
\usepackage{amssymb}
\usepackage{amsmath}
\usepackage{txfonts}
\usepackage{graphicx}
\usepackage{natbib}
\usepackage{rotating}
\usepackage{soul}
\usepackage{url}
\usepackage{epsfig}
\usepackage{longtable}
\usepackage{xcolor}
\definecolor{darkblue}{rgb}{0.0,0.1,0.6}
\definecolor{darkblue2}{rgb}{0.0,0.0,0.3}
\renewcommand{\hl}[1]{\textcolor{black}{#1}}
\definecolor{darkgreen}{rgb}{0.0,0.5,0.0}
\newcommand{\ch}[1]{\textcolor{black}{#1}}
\newcommand{\chn}[1]{\textcolor{black}{#1}}
\usepackage[utf8]{inputenc}
\usepackage{url}

\usepackage{comment}

\usepackage[colorlinks = true, linkcolor = blue, urlcolor  = blue, citecolor = blue, anchorcolor = blue]{hyperref}
\hypersetup{pdftitle={Triaxiality in Galaxy Clusters},
pdfauthor={S. Stapelberg}}

\usepackage{bbold}
\begin{document}

\title{Triaxiality in Galaxy Clusters: Mass versus Potential Reconstructions}
\titlerunning{Triaxiality in galaxy clusters}

\author{S. Stapelberg\inst{1,2}, C. Tchernin\inst{1,2},  D. Hug\inst{1,2}, E. T. Lau\inst{3} , M. Bartelmann\inst{1,2} }
\institute{\inst{1} Institut f\"ur Theoretische Physik, Heidelberg University, Philosophenweg 16, 69120 Heidelberg, Germany,
\\\inst{2} Institut f\"ur Theoretische Astrophysik, Zentrum f\"ur Astronomie, Heidelberg University, Philosophenweg 12, 69120 Heidelberg, Germany,
\\\inst{3} Department of Physics, University of Miami, Coral Gables, FL 33124, U.S.A}
\mail{stapelberg@uni-heidelberg.de }

\authorrunning{S. Stapelberg et al.}



\abstract
{Accounting for the triaxial shapes of galaxy clusters \hl{will become} important in the context of upcoming cosmological surveys. \chn{This will provide a challenge given that the density distribution of gas} cannot be described \ch{by} simple geometrical models without loss of information.}
{\chn{We show that, compared to the gas density distribution, the cluster gravitational potential can be better characterised by a simple 3D model and is more robust against fluctuations and ambiguities related to both masking of substructures and to shape definitions. Indeed, any perturbations in the gas density distribution can have a substantial influence on the derived thermodynamic properties, while the cluster potentials are smooth and can be approximated by a spheroidal model without as much loss of information.}}
{We \ch{use a statistical sample of 85 galaxy clusters from a large cosmological N-body + hydrodynamical simulation to investigate cluster shapes as a function of radius. In particular, we examine the shape of isodensity and isopotential \hl{shells} \ch{and analyze} how the \hl{results are affected by \ch{factors such as} the choice of component (gas vs. potential), the substructure removal algorithm (for the gas density) and the definition of the computation volume (interior vs. shells)}. }}
{\hl{We find that the orientation and axis ratio of the gas} isodensity contours are degenerate with the presence of substructures and unstable against \hl{fluctuations}. In addition, we observe that, as the derived cluster shape depends on the method used for removing the substructures, thermodynamic properties extracted from, for instance, the X-ray emissivity profile, suffer from this additional, and	often underestimated, bias. 
In contrast, \hl{for the smooth potential}, the shape reconstruction is \hl{largely unaffected by these factors} and converges towards simple geometrical models for both relaxed and dynamically active clusters. 
}
{
The \hl{observation that cluster potentials} can be represented \hl{better} by simple geometrical models and  reconstructed with a low level of systematics for both dynamically active  and relaxed clusters  (see our companion paper, Tchernin et al. 2020), suggests that by characterising galaxy clusters by their potential rather than by their mass, dynamically active and relaxed clusters could be combined in cosmological studies, improving statistics and lowering scatter. 
}

\maketitle
\section {Introduction}\label{sec:intro}

The statistics of the galaxy cluster population as a function of redshift and mass is often used to trace the  evolution  of   large-scale  structures	\cite[see e.g.][]{voit05, planck15_cosmocluster}. Such cosmological probes require well constrained cluster masses estimates to compare with the theoretical predictions. 
The X-ray bremsstrahlung emission \citep[e.g.][]{sarazin88, eckert18}, the signal from the thermal Sunyaev-Zel’dovich  effect \citep[tSZ, e.g.][]{SZ69,sayers11} and gravitational lensing measurements  \citep[e.g.][]{bartelmann10,jauzac19} are typical mass proxies, which are based on the assumption that the cluster matter distribution can be described by a simple spherical or spheroidal\footnote{A spheroid is an ellipsoid, whose axis ratio satisfy $a>(b=c)$, where $a$ is the major axis of the ellipsoid and $b,c$ the two minor axis.} model\footnote{While gravitational lensing measurements do not rely on symmetry assumptions, the cluster masses are often estimated using such assumptions.}. 
In addition, cosmological studies are often based on scaling relations derived from a sample of relaxed clusters, selected for being almost spherical and in hydrostatic equilibrium \citep{vikhlinin09,mantz16}. For instance, \citet{pratt19} review very well  the past and ongoing efforts to derive cosmological constraints using the sub-sample of relaxed clusters to tighten the existing mass-observables scaling relations (see their Sect.~4.5). These estimates are then compared to theoretical predictions that have been derived for these same simple geometrical models \citep[e.g. the spherical collapse model, \hl{see}][for some applications]{ps74,tozzi01,voit05}.

However, both simulations and observations have demonstrated for decades that the shape of isodensity contours in clusters is \hl{non-spherical and can deviate largely from spheroidicity, too (see e.g. }\citealt[][]{Limousin13} \hl{for a review).}
\hl{Cosmological simulations predict the formation of triaxial dark matter halos} \cite[e.g.][]{jing02} \hl{as a consequence of the anisotropic gravitational collapse for a Gaussian random field of primordial density fluctations.}
\chn{This also affects the relaxed clusters, whose shape can vary with radius \citep[e.g.][]{Dubinski91, allgood06,veraciro11,harvey19}}.
\hl{At the same time, cluster shapes depend on mass, with higher-mass halos being less spherical than lower-mass halos} \citep[e.g.][]{kasun05, allgood06, bett07}.
\chn{Depending on the selected cluster sample, different cosmological results may be obtained. Thus, accounting for the most precise cluster shape is crucial.}
\hl{Being the most massive bound objects, galaxy clusters formed latest in cosmic history and represent a population that is at least in parts out of dynamical equilibrium} \citep[e.g.][]{gao12, bartelmann13}.
\hl{Cosmological simulations suggest that the morphologies of cluster halos are closely linked to} the formation history of the galaxy clusters \citep{evrard93,suto16,lau20}. 
\hl{Their axial ratios are sensitive to $\sigma_8$ and (weakly) on $\Omega_m$ }\citep[see e.g.][and references therein]{buote12}
\hl{and expected to be closer to unity if dark matter is self-interacting} \citep[e.g.][]{dave01,peter13,banerjee20}.


\hl{In the context of upcoming cosmological studies, taking into account the three-dimensional cluster shapes and their influence \ch{on the reconstruction of cluster quantities} is going to play a critical role }\citep[e.g.][]{Limousin13}.
Observational mass biases arising from the symmetry assumption are already well documented \citep[see e.g.][]{rasia12}. 
Due to the dependence of the X-ray emissivity  on the squared gas density and on the steep slope of gas density, the cluster morphology derived from X-rays appears typically more spherical than it would be if derived from other observables \citep[combining different observational probes is necessary to constrain the cluster shape e.g,][]{Morandi10, morandi12, Limousin13, sereno18}. \ch{In addition, the presence of substructures \hl{may lead to underestimation of} the total hydrostatic mass (HE mass) derived from X-rays \citep[e.g.][]{ettori13}.}
A dedicated analysis of the effect of the assumed symmetry on the results obtained from hydrostatic equilibrium can be found in \citet[][]{chiu12}. 
Regarding \hl{gravitational lensing measurements, cluster masses derived under the assumption of spherical symmetry may be over- or underestimated} depending \hl{on} the orientation of the principal axis of \ch{the lens mass} distribution with respect to	the line-of-sight \citep[][]{meneghetti10,osato18,lee18}.



Contrary to most of the studies listed above and to the analysis performed in \citet[][]{buote12} -- where the authors computed the orientation-average biases arising when masses of ellipsoidal clusters are estimated assuming spherical symmetry -- in the present study, we do not assume that the density distribution is stratified in ellipsoids of constant axis ratio and orientation, as this is clearly not the case. Instead we are interested in the variation of the shape of \ch{isodensity layers with cluster-centric distance.} \hl{We investigate the} impact of \hl{asymmetries, fluctuations} and substructure-removal methods on the shape of these \ch{isodensity layers}. 
Substructures are clumps of matter \hl{that are} not yet in virial equilibrium with the cluster \hl{gravitational potential}. This implies that their temperature, metallicity and density are not representative for the surrounding intracluster medium (ICM), and therefore, that they should be removed from \ch{the analysis if we wish to derive bulk} properties such as the hydrostatic mass. This  step is particularly important for the treatment of X-ray observations, as the signal emitted by these density inhomogeneities is enhanced relatively to the signal from the ICM gas \citep[e.g.][]{ettori13}.


Modelling the 3D morphology of a cluster from 2D observations is complicated due to projection effects \hl{arising from the unknown orientation angle and the intrinsic ellipticity of the cluster, but also from} the contribution of the substructures. We show here that this task \ch{is not trivial due to ambiguities even} \chn{if we have at hand} the full	3D density distribution of simulated clusters. \ch{For instance}, substructure removal is performed for an assumed cluster morphology and this, in turn, \hl{biases the resulting density distribution -- as can be observed when using} the azimuthal median method to remove substructures \citep[][]{eckert15}: in each spherical shell, the median is used instead of the mean. While being very efficient for spherically symmetric clusters \citep{zhuravleva13}, we saw in our companion study \citep{tchernin20} that applying this method to an elliptical cluster biases low the resulting profile (even for a cluster with small ellipticity, see Fig. A.8 of \hl{that} study). However, \hl{there} exist also less strict selection criteria to identify \ch{substructures in spherical shells, such as} the $X$-$\sigma$ criterion,  where $X$ corresponds to a density threshold above which the	pixels	with larger density  values  are excluded from the analysis \citep[\ch{for instance $X=3.5$ is used in}][]{zhuravleva13, lau15}. 

\hl{\chn{Here} we} 
investigate the three-dimensional shapes of \chn{a statistical sample of simulated galaxy clusters drawn from a cosmological hydrodynamic simulation}.
\chn{This includes a “reference sample” of 32 relaxed clusters, for which we explore several parameters of our shape determination methods, and an additional sample of 53 non-relaxed clusters, for which we measure shapes with the best working set of parameters.}
\hl{We show} that removing substructures for an assumed spherical symmetry is significantly affecting the shape of the gas density distribution and should therefore be carefully taken into account in the systematics of the extracted \hl{mass} profiles.
In particular, in Sect.~\ref{sec:results_relaxed} we show that the analysis of the principal axes of the gas density distribution \hl{is heavily affected by} the presence of \hl{fluctuations and \chn{by} the specific choice of substructure removal technique, even for relaxed clusters}. 
We demonstrate that 
the \hl{distribution of the gravitational potential, in contrast}, can be described by a \ch{simpler} morphology and that \hl{fitting the} potential isocontours is \hl{much more robust against \chn{the presence of local fluctuations and perturbations. 
We also note that differences in methodical details, such as the definition of the integration domain, have a smaller impact on the potential due to its simple geometry.}
\hl{The gravitational potential contains all relevant information on the cluster morphology and can be related straightforwardly to a large number of multi-wavelength observables including X-ray, SZ, gravitational lensing and galaxy kinematics, which allow for a joint reconstruction of cluster potentials \citep{konrad13,sarli14,tchernin15,stock2015,majer16}. }
We discuss our results and their implications in Sect.~\ref{sec:disc} and conclude with the advantages of using the gravitational potential in Sect.~\ref{sec:conc}.}

\section{Simulation and methods}

\label{sec:ana}

\subsection{Cosmological simulations} \label{sec:omega500_sim}

\begin{figure}
  \centering
  \includegraphics[width=\columnwidth,angle=0]{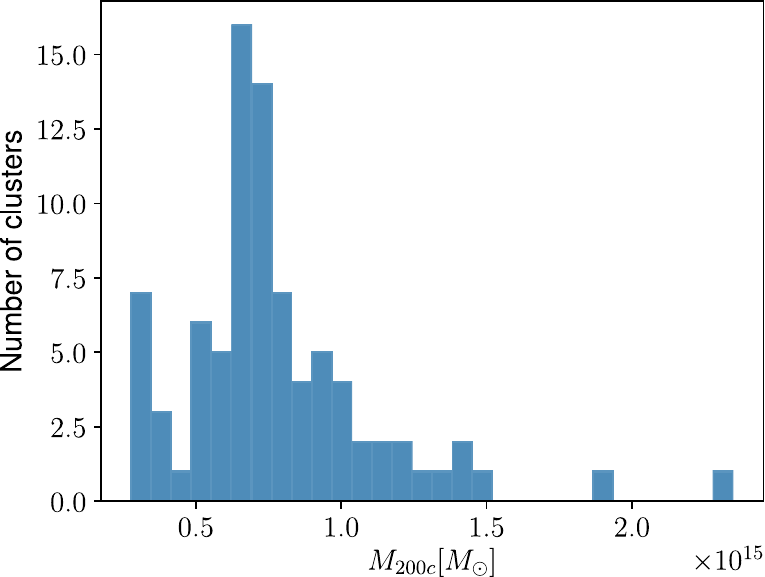}
\caption{\hl{Distribution of virial masses for the 85 simulated clusters of the \textsc{Omega500} sample used in this work.}} 
\label{fig:cluster_hist}
\end{figure}

\hl{\chn{In this study,} we use \chn{a statistical sample of 85}} galaxy clusters 
%
%
\hl{extracted from} the non-radiative \chn{hydrodynamic cosmological simulation} {\textsc{Omega500}} \citep{nelson14}, \hl{which was specially designed for studying galaxy clusters and} also used in the companion paper \citep{tchernin20}. \hl{\textsc{Omega500} was} performed using the \hl{Eulerian} \textsc{Adaptive Refinement Tree} (ART) N-body+gas-dynamics code \citep{kravtsov99, rudd08}, 
\hl{adopting a flat $\Lambda$CDM cosmology with $H_0 = 70 \, \mathrm{km} \, \mathrm{s}^{-1} \, \mathrm{Mpc}^{-1}$, $\Omega_m = 0.27$, $\Omega_{\Lambda} =0.73$, $\Omega_{b} = 0.0469$ and $\sigma_8 = 0.82$.  The evolution of dark matter and gas was followed on an adaptive grid, with a co-moving box size length of $500 \, \mathrm{Mpc}/h$ resolved using a uniform $512^3$ root grid with  $8$ levels of mesh refinement.}
\hl{Clusters were identified using a halo finder that detects mass peaks on a given scale and then performs an iterative search for the center of mass \citep[see][for details]{nelson12}. The full sample of clusters consists of 85 objects \chn{at $z = 0$} with masses $M_{200c} \gtrsim 1.94 \times 10^{14} h^{-1} M_{\odot} $, shown in a histogram in Fig.~\ref{fig:cluster_hist}.}

\hl{We divide this cluster sample into `relaxed' and `non-relaxed' clusters based 
on 
the structural morphology of their X-ray images \hl{according to} the `symmetry$-$peakiness$-$alignment' (SPA) criterion by} \citep[][]{mantz15}. \hl{This criterion applies cuts in three shape parameters referring to the regularity of X-ray isophotes; and was applied to mock \emph{Chandra} maps of the \textsc{Omega500} clusters previously} \citep{shi16b, tchernin20}.
\hl{This classification results in 32 relaxed and 53 unrelaxed clusters. Its purpose is to study the impact of the clusters' dynamical state, which is expected to correlate with the deviation \chn{from sphericity}.
For both categories of clusters, we will carry out the same analysis and comparisons. 
}

\begin{figure}
  \centering
  \includegraphics[width=\columnwidth,angle=0]{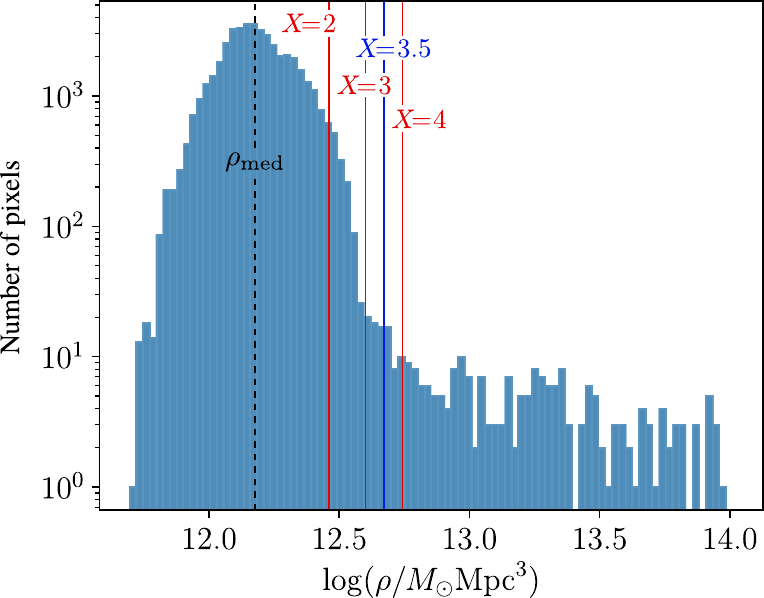}
\caption{\hl{Illustration of the $X$-sigma method. Shown are gas density values of the relaxed cluster CL1 in an arbitrary, thin spherical shell of width $\Delta r \approx 11 \mathrm{kpc}$ between $r_{500}$ and $r_{200}$ \chn{(this is an example choice for illustration only)}
The dashed line shows the median density, while the blue solid line shows the threshold for $X = 3.5$. The red solid lines show further, arbitrary example choices $(X = 2, 3, 4)$ to give an impression of how much variations of order unity in $X$ affect the distribution. 
}
} 
\label{fig:cluster_ind_hist}
\end{figure}

\hl{In our previous study, we derived} the systematics of the potential and HE mass for the \textsc{Omega500} clusters, assuming spherical symmetry \citep{tchernin20}. We observed that the reconstructed potential is less affected by this \chn{simplifying assumption} than the HE mass. We will now go one step further and study the cluster morphology using the \chn{entire} gas density distribution. We will show in Sect.~\ref{sec:results_relaxed} that the \hl{gas density} distribution $\rho_g(\boldsymbol x)$ in such galaxy clusters cannot be approximated with simple geometrical models \hl{and is significantly affected by the presence of substructures and fluctuations in the bulk gas density.} We will then compare these results to the cluster morphological information extracted using the potential distribution $\Phi(\boldsymbol x)$.


\subsection {Substructure removal}\label{sec:subremoval}
The triaxiality of clusters is degenerate with the presence of substructures.
\hl{Substructures are gas-rich subhalos of matter that are not yet in virial \chn{or} hydrostatic equilibrium with the gravitational potential.
The impact of these substructures must be taken into account when estimating the profiles or three dimensional shapes of clusters based on observables related to the gas density distribution, such as X-ray emissivity \citep[e.g.][]{lau15}.
In particular, to avoid large systematic biases for instance in the ellipticity of cluster shapes, the contributions caused by inhomogeneities need to be \chn{removed} from the signal of the bulk gas component. 
\chn{The bulk properties are in fact those relevant for characterising galaxy clusters -- whether by their mass or their gravitational potential -- under the assumption of hydrostatic equilibrium.} 
}

\begin{figure*}
  \centering
  \includegraphics[width=0.85\textwidth,angle=0]{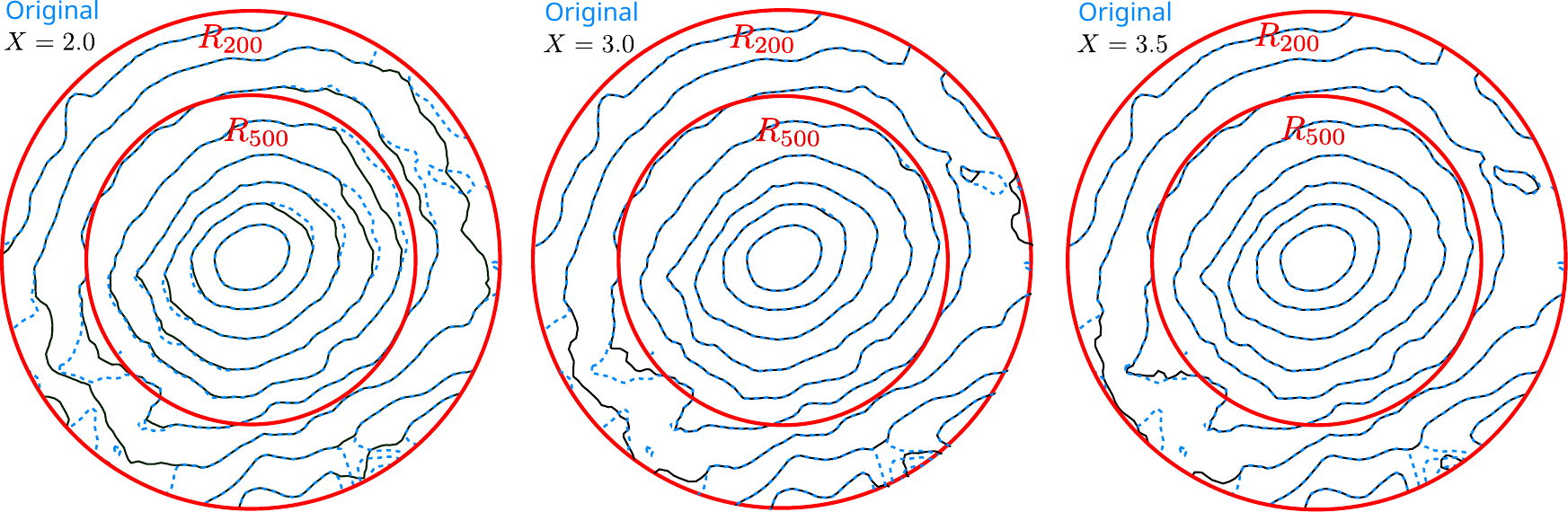}
\caption{Illustration of the substructure--triaxiality degeneracy: The figure shows 2D projected gas density isocontours of the simulated, average-mass relaxed cluster CL85, \chn{which we have picked here because its} major axis is almost perpendicular to the viewing direction.
Black: Density isocontours after substructure removal for different choices.
Blue dashed: Original isocontours. 
While $X = 3.5$ does not remove all substructures, the stricter choice $X= 2$ also removes ellipticity from contours in the bulk within $r_{500}$.
}
\label{fig:2sigma}
\end{figure*}

In \citet{tchernin20}, we \hl{considered} two methods to remove \hl{the impact of} substructures, viz. the X-$\sigma$ criteri\hl{on} and the median method.
\hl{While the median method replaces the density distribution by its median in spherical shells, the X-$\sigma$ criterion \citep{zhuravleva13} preserves the three dimensional cluster shape information by using the shape of the logarithmic probability density distribution. 
The density distribution in spherical shells can be described by a log-normal function, corresponding to a nearly hydrostatic bulk component, and a high-density tail caused by substructure-induced inhomogeneities. 
For illustration, an example for this is shown for the relaxed cluster CL1 in Fig.~\ref{fig:cluster_ind_hist}. 
%
The $X$-$\sigma$ method removes gas cells with a logarithmic density of a factor $X$ times $\sigma$ above the median, which are assigned to the high-density tail:}
\begin{align}
	\hl{ \log_{10} \rho_g} \; &{>} \; \hl{ \log_{10} \{ \rho_g \} + X \sigma } 
\end{align}
\hl{Here, $\sigma$ is the $\log_ {10}$-based standard deviation of the density distribution.}
%
\hl{The specific choice for $X$ controls how \chn{sharp the} \ch{cut will be}. A typical empirically motivated value used in the literature is $X=3.5$ \citep{zhuravleva13, lau15}. However, slightly higher or lower values might be equally viable, given the unsharp \ch{transition} between the log-normal component and the high-density tail. A few example values are shown in Fig.~\ref{fig:cluster_ind_hist}.
%
}
\hl{\ch{\chn{For the computation of shape parameters, we substitute} the holes left by masking substructures \chn{in two different ways by} substituting} (a) the median \chn{gas density at that cluster-centric radius, as was done by \citep{chen19}};
and (b) the threshold density value. While the former \chn{choice is representative for the distribution of gas density values} within spherical shells, the latter preserves continuity and intrinsic asymmetries of the cluster at the expense of possibly overestimated gas density values at the positions of substructures.
\ch{We will apply and test the outcomes of both methods throughout this work.}
}


\hl{\ch{For the cut parameter $X$, aside from } the aforementioned \chn{reference} value of $X = 3.5$, we \ch{apply also a} slightly stricter and \ch{a} less strict threshold $X \in \{3, 4 \}$ to study the impact of $X$. \ch{In addition, we} include \ch{the} choice $X = 2.0$, for which we saw in \cite{tchernin20} that this produces a mean density profile more similar to that of the median method.}
\hl{We thus} obtain four distinct “cleaned” simulation sets, which we will call “2S”, “3S”, “3.5S” \chn{(or “reference”)} and “4S”, respectively.
\hl{
For each of these \ch{data sets, we} apply and study both \ch{filling} methods (a) and (b) \ch{to} compare their impact on the 3D cluster morphologies.
For comparisons with the gravitational potential distribution, we will furthermore distinguish between two different representations of the potential data set: (i) the \chn{“raw” non-processed} potential of the cluster \ch{including all subclump} contributions; \ch{and } (ii) a version in which the regions identified to contain substructures are \ch{replaced by the median potential in the same fashion as for the gas density filling method (a).} 
This latter version \ch{puts emphasis on methodical consistency while} the former \ch{comparison (i.e. between substructure-cleaned gas density and raw potential)} \chn{is interesting because it contrasts the optimum we can hope to achieve with the gas density (in terms of a smooth estimator of the intrinsic shape) after pre-processing it with the result we can expect from the non-processed potential.}
}

%

\hl{We note that the prevalence of substructures is expected to be highest \ch{in the outer regions, e.g. close to $r_{500}$, which are dynamically younger and still show ongoing accretion of matter \citep[e.g.][]{nagai11,roncarelli13, vazza13}.}
Nonetheless, applying substructure removal techniques can introduce uncertainties in the definition of cluster shapes at any radii. \chn{Moreover,} there is no sharp distinction between inhomogeneities due to non-virialised gas clumps and \ch{intrinsic} asymmetries of the bulk gas component.  
\ch{As a result,} substructures and triaxiality can be highly degenerate (c.f. Fig.~\ref{fig:2sigma}); and the removal of substructures, which is always required for the gas density but may not be required for the potential, may artificially bias cluster shapes towards sphericity.}
\hl{Using all the different aforementioned data sets, we will} investigate the robustness \hl{of cluster shapes with respect to the value of $X$, the influence of removing or replacing the pixels, and the sensitivity to residual density fluctuations\footnote{\chn{These are fluctuations not identified and removed by the $X$-$\sigma$ criterion either because their amplitude is too weak, or they belong to the small-scale fluctuations intrinsically present in the bulk component e.g. due to gas motions \citep[see e.g.][]{zhuravleva13}.}} in the “bulk” gas component. }

\subsection{Cluster shape definitions} \label{sec:shape}

We \hl{now} describe the methods \ch{used} to analyze the shape of the simulated clusters.
\hl{An ellipsoidal body can be \ch{defined} by its three principal axes $a \geq b \geq c$, \ch{while the orientation of the axis of longest elongation will be denoted by \chn{an angle vector}} $\phi$. For characterising the shape itself, only two axis ratios are required. 
For a general mass or potential distribution $\rho$, the principal axis vectors are related to the eigenvectors of the mass tensor\footnote{The definition of $\boldsymbol M$ is \hl{analog} to the definition of the covariance matrix for a distribution of data points.}
}
\begin{align}\hl{\mathbf M = \int_V \rho(\boldsymbol x) \boldsymbol x \boldsymbol x^{T} \, \mathrm d V, } \label{eq:pca} \end{align}
\hl{for a given volume $V$, where $\boldsymbol x$ is the position vector with respect to the centroid of the mass distribution.}
\hl{For the discrete data cubes provided by the simulations, the integral in Eq.~\ref{eq:pca} turns into a sum over grid pixels. This sum can be computed numerically and the resulting matrix $M_{ij}$ can be diagonalised to obtain its eigenvectors and eigenvalues.} 


\hl{In order to characterise the cluster shape, we define the ellipticity $\epsilon$ and the aspheroidicity $A$ as}
\begin{align}
	\hl{ \epsilon} &= \hl{1 - \frac{c^2}{a^2} } \label{eq:eps} \\
	\hl{ A } &= \hl{ 1 - \frac{c^2}{b^2} }, \label{eq:asph} 
\end{align}
\hl{where $a \geq b \geq c$ are the three principal axis lengths that correspond to the three \chn{eigenvalues $\lambda_1, \lambda_2, \lambda_3$} of the mass tensor,  $a \propto \sqrt{\lambda_1}$, $b \propto \sqrt{\lambda_2}$, $c \propto \sqrt{\lambda_3}$.
Note that $b = c$ describes the \ch{case} of a spheroid.}
\ch{\chn{The ellipsoidal parametrization of shapes for the cluster gas and potential distributions is motivated} by earlier theoretical studies suggesting that the isopotential surfaces of dark matter halos are well approximated by triaxial ellipsoids even for mass distributions that are ellipsoidal themselves \citep[e.g.][]{lee2003, hayashi2007}.}


\begin{figure}
  \centering
  \includegraphics[width=\columnwidth,angle=0]{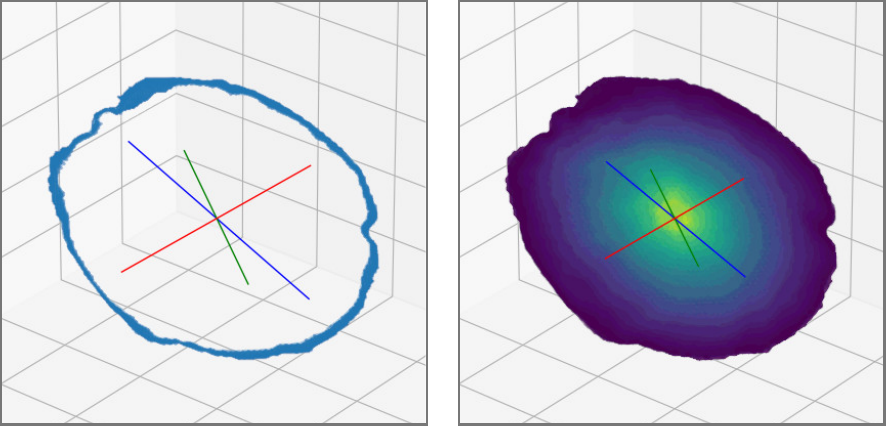}
\caption{Illustration of the \hl{“local”} and \hl{“inner” choice of PCA integration domain} (Eq.~\ref{eq:pca}), \hl{for the simulated relaxed cluster CL6}. Left: \hl{Slicing through a thin gas density layer, right: Slicing through the density distribution enclosed by the same layer. The three lines illustrate the directions of the three principal axes.} } 
\label{fig:methods3DPCA}
\end{figure}

\hl{The morphology of a cluster \ch{is} generally \ch{expected to evolve} with the distance from the cluster center \citep[e.g.][]{allgood06}. 
\ch{This} radial \ch{dependence of the cluster shape} may contain potentially interesting information about the history of cluster formation, \ch{as the geometry of different layers is sensitive to cosmology and structure growth at different epochs }\citep[e.g.][]{lau20}.}
\hl{\ch{In this work we} study the cluster shape evolution as a function of radius. \ch{This is done by computing} the mass tensor for different `density layers' (or potential layers) of each cluster \ch{individually -- to obtain} profiles of \ch{the} shape parameters of interest, using the generalised radius $r$ defined in Eq.~\ref{eq:rad}. We start by binning the gas and potential distributions into \chn{isodensity (isopotential)} shells of constant width in logarithmic space.
The width of each bin is set to $\Delta (\ln \rho) = \ln (\rho_{i+1}/{\rho_i}) = 0.05$ for the gas density and $\Delta (\ln \rho) = 0.02$ for the potential.
These values are \ch{arbitrary choices} motivated by the requirement that the resulting shells be both large enough compared to the pixel size given by the finite resolution of the simulation, and small enough to still obtain smooth radial profiles.
We have tested that small variations in the bin width do not influence our results and conclusions (see also Sect.~\ref{methods_and_datasets}, \ch{where we comment further on this}).}

\hl{For each bin, we apply two different methods for determining the shape of the cluster, which differ from each other by computing \ch{the integral in} $M_{ij}$ \ch{over} different \ch{domains}: 
(1) \ch{over the corresponding} thin shell only ($\rho = \mathrm{const.}$) to define a \emph{“local”} shape measure; (2) over the entire \emph{“inner”} \ch{spatial} volume enclosed by the \ch{shell (including the shell itself)}, \ch{such that \chn{each bin effectively acts as an overdensity threshold defining a region in which the shape of the enclosed cluster mass is measured. }}
An illustration of the two approaches is given in Fig.~\ref{fig:methods3DPCA}.}
For brevity, we will refer to these two variants respectively as \emph{local} and \emph{inner} method hereafter. 
\hl{We note that the inner method may produce a slightly different result at a given radius compared to the local method, because the mass tensor is computed from a larger region, picking up contributions from the interior mass distribution and thus from different layers.
This may cause its shape evolution in the triaxial profiles to \ch{depend also on the density profile and to} `lag' in radius compared to the measurement in the local domain, \ch{as pointed out in a different context by \citep{zemp11}}.
On the other hand, the inner method is expected to be smoother and less affected by the radial fluctuations of each isodensity layer.
}

\hl{In order to derive triaxial profiles of the shape parameters and the cluster orientation, we introduce the generalised ellipsoidal radius,} which is updated with the information on the cluster shape \hl{as defined by} \cite{Dubinski91}:
\begin{align} \label{eq:rad}
    r = \sqrt{ \tilde x^2 + \left( \frac{\tilde y}{q} \right)^2 + \left( \frac{\tilde z }{s} \right)^2 },
\end{align}
where $q := b/a$ and $s = c/a \le q$ are the principal axes ratios and the tilde denotes coordinates rotated into the principal-axis frame of the mass tensor. \hl{Note that for $a = b = c$, this definition} reduces to the concentric radius $\smash{r = \sqrt{x^2+y^2+z^2}}$.
All profiles shown in this paper are displayed as a function of the \hl{generalized ellipsoidal radius. \chn{In addition to the quantities we introduced so far, we define a parameter 
\begin{align}
	\Delta \phi(r) := \angle ( \phi(r), \bar \phi ) \label{eq:delta_phi} 
\end{align}
specifying the relative angle between the major axis at the current radius $r$ and the medan major axis orientation of the cluster as a “fiducial” axis. This allows to quantify the radial variation of orientation in a direction- and coordinate-independent fashion, adapting to the clusters' own average principal frames. By quantifying and comparing the change in orientation in the principal axis between different methods or data sets, we can investigate how sensitive the latter are to fluctuations in the isodensity or isopotential shells perturbing the principal axis directions.} }

\section{Results: 3D morphologies of galaxy clusters} \label{sec:results_relaxed}

\hl{We apply the PCA methods defined in Sect.~\ref{sec:ana} to the gas density and potential distributions of the \ch{85} \textsc{Omega500} clusters. 
For the gas density, we \ch{study} the 2S, 3S, 3.5S and 4S data sets in the variants (a) and (b) discussed in Sect.~\ref{sec:shape}. For the potential, we will distinguish between variants (i) and (ii) discussed in Sect.~\ref{sec:subremoval}. 
For each data set, we compute the triaxial ellipsoidal profiles of the quantities $\epsilon, A$ and  $\Delta \phi $ \chn{(see Eqs.~\ref{eq:eps},~\ref{eq:asph},~and~\ref{eq:delta_phi})} averaged over all clusters of the \ch{relaxed and non-relaxed cluster sub-samples.}
These profiles are normalised by the halo virial \ch{radii $r_{200}$ and then represented
by the median among the cluster population\footnote{Note that we use the median here instead of the mean because it is less sensitive to outliers e.g. caused by largely perturbed individual clusters.}. We also compute the quartiles of the sample to measure the scatter of each quantity.
}
We start by focusing on relaxed clusters as our main sample of study in Sect.~\ref{sec:relaxed_gas_results}. We then extend our analysis to the non-relaxed cluster population in the subsequent Section~\ref{sec:results_unrelaxed}.
}





\subsection{Cluster shapes from relaxed galaxy clusters} \label{sec:relaxed_gas_results}

\hl{
We \chn{begin} our analysis by determining the ellipticity $\epsilon(r)$, the aspheroidicity $A(r)$ and the major axis deviation $\Delta \phi(r) := \angle(\phi(r), \bar \phi) $ of the gas density layers, as defined in Sect.~\ref{sec:shape}, for the different datasets defined in Sect.~\ref{sec:subremoval}. 
}
\hl{We begin by performing \ch{our analysis for the} \chn{reference} configuration 3.5S \ch{in the “threshold” filling method b.}  
For this data set, we investigate the cluster shape parameters and their radial behavior for the gas density both for the inner and the local integration domains. 
We then compare the results for variations in the parameters of our methods, to study how the outcome depends on these parameters. In particular, we compare the results for 2S, 3S, 4S for methods \chn{(a) and (b)}, and when using different domains (inner vs. local). This allows to probe sources of uncertainty but also to study the impact of substructures and their removal.}


\subsubsection{ Cluster shapes from the gas density for the default substructure method (3.5S) } \label{sec:relaxed_gas_domain}

\begin{figure*}[t!]
\begin{subfigure}[b]{0.328\textwidth}
  \centering
  \includegraphics[width=1.01\columnwidth,angle=0,clip]{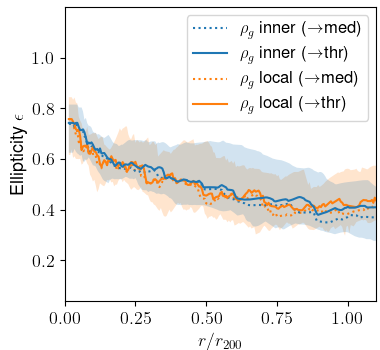}
  \caption{} 
   \label{fig:relaxed_gas_domain1}
\end{subfigure}
\begin{subfigure}[b]{0.328\textwidth}
  \centering
  \includegraphics[width=1.01\columnwidth,angle=0,clip]{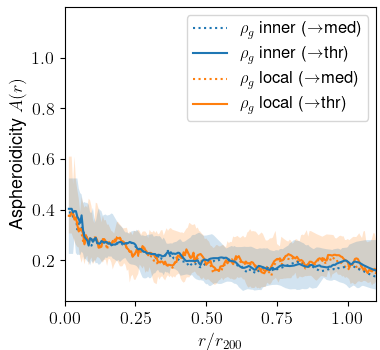}
  \caption{}
   \label{fig:relaxed_gas_domain2}
\end{subfigure}
\hspace{0.5pt}
\begin{subfigure}[b]{0.328\textwidth}
  \centering
  \includegraphics[width=1.01\columnwidth,angle=0,clip]{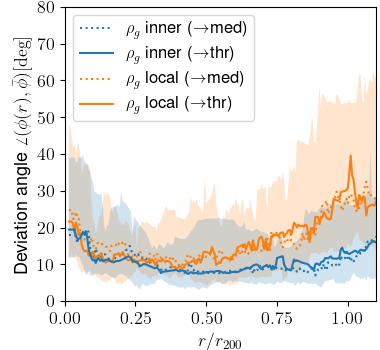}
  \caption{}
   \label{fig:relaxed_gas_domain3}
\end{subfigure}
\caption{\hl{Morphological parameters \chn{$\epsilon(r)$ (Eq.~\ref{eq:eps}), $A(r)$ (Eq.~\ref{eq:asph})} and deviation in orientation $\Delta \phi$ (Eq.~\ref{eq:delta_phi}) from the gas density (3.5S) for relaxed clusters. The semi-transparent areas enclose the quartiles of the cluster sample, while “med” and “thr” denote whether regions occupied by substructures are replaced by the median or threshold approach (“a” and “b” defined in Sect.~\ref{sec:subremoval}).} }
	\label{fig:relaxed_gas_sub}
\end{figure*}

\begin{figure*}[t!]
\begin{subfigure}[b]{0.328\textwidth}
  \centering
  \includegraphics[width=\columnwidth,angle=0]{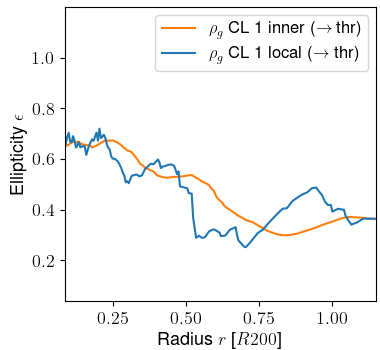}
  \caption{}
   \label{fig:relaxed_gas_indiv1}
\end{subfigure}
\hfill
\begin{subfigure}[b]{0.328\textwidth}
  \centering
  \includegraphics[width=\columnwidth,angle=0]{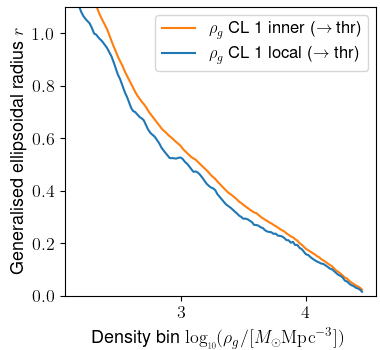}
\caption{
}
   \label{fig:relaxed_gas_indiv2}
\end{subfigure}
\hfill
\begin{subfigure}[b]{0.328\textwidth}
  \centering
  \includegraphics[width=\columnwidth,angle=0]{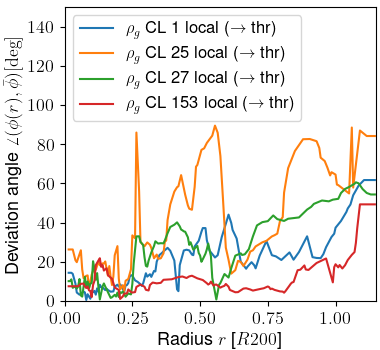}
  \caption{}
   \label{fig:relaxed_gas_indiv3}
\end{subfigure}

\caption{\hl{Individual cluster result examples \chn{for $\epsilon(r)$, $r(\log_{10} \rho)$, and $\Delta \phi(r)$} for the gas density. \chn{The left panel shows a comparison between inner and local domain for CL1 to illustrate the lagging effect mentioned in Sect.~\ref{sec:relaxed_gas_domain}. The middle panel shows the fluctuation behavior of the generalised ellipsoidal radius for the same cluster. The right panel compares the variation in major axis orientation for various different clusters in one particular configuration (local) to highlight the individual fluctuation behavior in orientation and its differences between the clusters. } These figures are intended for illustration only, complementary to our statistical analysis using the full cluster sample. The clusters selected for these figures were picked using a random number generator. 
\hl{The labels “med” and “thr” denote whether regions occupied by substructures are replaced by the median or threshold approach (“a” and “b” defined in Sect.~\ref{sec:subremoval}).}}}
   \label{fig:relaxed_gas_indiv}

\end{figure*}

\begin{figure*}[t!]

\begin{subfigure}[b]{0.328\textwidth}
  \centering
  \includegraphics[width=1.01\columnwidth,angle=0,clip]{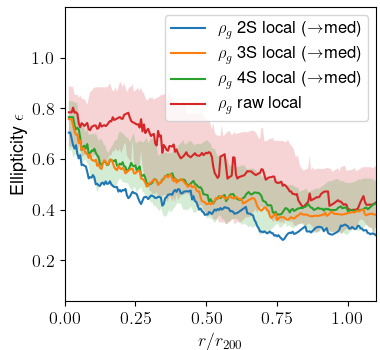}
  \caption{}
   \label{fig:relaxed_gas_domain5}
\end{subfigure}
\begin{subfigure}[b]{0.328\textwidth}
  \centering
  \includegraphics[width=1.01\columnwidth,angle=0,clip]{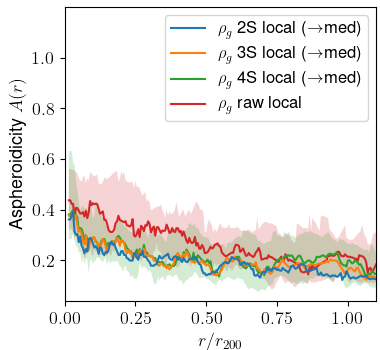}
  \caption{}
   \label{fig:relaxed_gas_domain6}
\end{subfigure}
\hspace{0.5pt}
\begin{subfigure}[b]{0.328\textwidth}
  \centering
  \includegraphics[width=1.01\columnwidth,angle=0,clip]{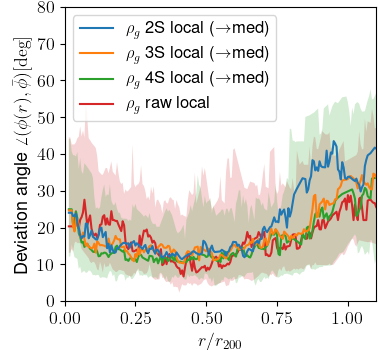}
  \caption{}
   \label{fig:relaxed_gas_domain7}
\end{subfigure}

\label{fig:relaxed_gas_diff_sub}
\caption{\hl{Comparison between substructure-affected (raw) and various cleaned datasets using different cut parameters (2S, 3S and 4S) for a fixed configuration (e.g. local domain + median method). \chn{The semi-transparent areas enclose the quartiles of the cluster sample -- here shown only for the raw and the 4S results to avoid a crowded plot. As before, “med” denotes that regions occupied by substructures are replaced by the median approach (see Sect.~\ref{sec:subremoval}). The label “raw” denotes the non-processed original dataset.}}}
\end{figure*}

\hl{For our \chn{reference} dataset 3.5S, we determine the median cluster shape parameters $\epsilon, A$ and $\Delta \phi$ and their quartiles, \ch{which yields the followig results for the inner and local domains:}}

\begin{itemize}
	\item \hl{Ellipticity:\\
	\ch{The ellipticity} $\epsilon$ of the gas density layers is shown in Fig.~\ref{fig:relaxed_gas_domain1}. 
	\ch{Its median decreases} from values of $\epsilon \sim 0.75$ near the center \ch{down to} ca. $\epsilon \sim 0.4$ towards the outskirts -- regardless of the PCA domain. \ch{Around $r \sim r_{200}$, the ellipticity $\epsilon$ slightly increases again}. The quartiles deviate \ch{typically} by ca. 20 to 25 percent (in peaks up to ca. 50 percent). \ch{The upper quartile shows a larger deviation from the median than the lower quartile.}
For the inner domain, \ch{as expected, the features of the curve are smoothed out and slightly shifted to higher $r$} due to the delayed response to radial variations that we described in Sect.~\ref{sec:shape}. \ch{This “lagging” effect can be seen even more clearly when plotting the results for single clusters, as shown for illustration in \chn{the example of} Fig.~\ref{fig:relaxed_gas_indiv1}.} }\\


	\item \hl{
	Aspheroidicity:\\
	\ch{The aspheroidicity profile $A(r)$ is} shown in Fig.~\ref{fig:relaxed_gas_domain2}. \ch{Here, we can} note that median and quartiles drop by ca. 60 percent from values around $0.4$ (median), \ch{$0.2$ (lower quartile), and $0.6$ (upper quartile)} within the first 10\% of $r_{200}$. This is followed by a nearly constant phase in which the median shows a near plateau until $r = 0.2 r_{200}$ followed by a slight decrease plus fluctuations. The median stays above a value of $0.15$ throughout, whereas the quartiles deviate by \ch{values around $\sim 0.1$ from the median and up to $\sim 0.2$ in the peaks (both inner and local)}.
	The result for the inner PCA domain differs from that of the local domain mainly by a slightly smoother behavior, the aforementioned lagging effect and lower values of the upper quartiles. \\
}

   \item{} \hl{Relative orientation:\\ The relative angle $\Delta \phi = \angle (\phi(r), \bar \phi)$ between the major axis at radius $r$ \ch{and the median cluster major axis, shown} in Fig.~\ref{fig:relaxed_gas_domain3}, \ch{fluctuates very strongly between $5^{\circ}$ and $30^{\circ}$ \chn{in the quartiles} (median: $\sim 11^{\circ}$) between \chn{$0 < r/r_{200} \lesssim 0.5 $} and increases by up to a factor two towards the virial radius for the local domain. For the inner domain, the orientation is more stable, fluctuating around $10\%$ and increasing less \chn{steeply} with radius as a result of the lagging effect.
   The quartiles, however, still show a large spread, with an inter-quartile distance between $10^{\circ}$ and $20^{\circ}$ typically.} }
   
\hl{A look at the same diagrams for \ch{examples of randomly picked} individual clusters, \ch{e.g. Fig.~\ref{fig:relaxed_gas_indiv3}}, shows \ch{that radial} fluctuations \ch{between layers are typically high, with rapid jumps by several ten degrees occuring between some layers. Investigation of the individual gas density data shows that these rapid changes in orientation are caused by intrinsic fluctuations in the density.
For instance, for CL25 shown in Fig.~\ref{fig:relaxed_gas_indiv3}, residual fluctuations perturb the shape of isocontours, which are rather close to sphericity at several radii (e.g. $\epsilon_{\rho_g} \sim 0.2$ to $0.4$ for this cluster in the outskirts) such that the orientation of the major axis becomes unstable.
On top of these fluctuations, as we see in Fig.~\ref{fig:relaxed_gas_indiv3}}, there is a monotonic trend of increasing deviation angle with radius. We find that this is a property shared by many clusters and resulting from perturbations and asymmetries in the gas density distribution increasing with radius. An example for this is CL27, for which the isocontours become elongated and seemingly disturbed in the outskirts.
} \ch{In sum, these various radial variations in the orientations of clusters \chn{contribute substantially to} the size of the inter-quartile range of Fig.~\ref{fig:relaxed_gas_domain3}.}

\end{itemize}





\subsubsection{Cluster shapes from the gas density for different data sets: Raw, 2S, 3S, 4S} \label{sec:relaxed_gas_different_sub}

\hl{We now extend our analysis to data sets produced for different substructure removal methods in order to explore how sensitive the derived cluster shape parameters are to the \ch{cut-off} threshold $X$. 
\ch{In order to show how important it is to mask substructures in the first place, we also include the original non-processed (“raw”) density distribution in our comparison. \chn{We note that the impact of substructures is even more important when deriving shape information from X-ray data, given the squared dependence of the emissivity on the density}.}
In the following, we present the outcome for these different data sets, which is shown in Figs.~\ref{fig:relaxed_gas_domain5} to~\ref{fig:relaxed_gas_domain6}, and compare to the \chn{reference} data set:}


\begin{itemize}
	\item \hl{\ch{Raw} data set:\\\chn{If the spurious signal from non-virialised substructures is not removed}, the axis \ch{lengths deviate considerably from each other -- e.g. for the ellipticity and the aspheroidicity, the median curves deviate from each other by ca. 20 to 30\% below $r \sim 0.75 r_{200}$, while the deviation is smaller at large radii. The inter-quartile range increases dramatically, e.g. enclosing values from $\epsilon \sim 0.4$ to $\epsilon \sim 0.9$ for the ellipticity.
	In contrast, the fluctuations in orientation show opposite trends at small and large radii: Up to $r/r_{200} \lesssim 0.4$, the raw, non-processed data leads to slightly higher fluctuations than the substructure-cleaned data sets, as can be seen from the median and the upper quartiles. 
	In the outskirts, $r/r_{200} \gtrsim 0.75$, however, the trend is opposite: Here, the median and the upper quartile are similar or lower than for the substructure-cleaned data sets.
	This suggests that for lower and intermediate radii, the substructures \chn{identified} by the $X$-$\sigma$ criterion introduce significant deviations from the median cluster orientation; while at large radii, substructure-removal on the opposite increases the deviation. 
	The median orientation itself differs little between the raw and substructure-cleaned data sets, typically by orders of a few degree only. 
	Thus, the visible deviations between the data sets are mostly showing local deviations between the major axis orientations.
	In the outskirts, these deviations are the result of a combination of low ellipticity of cluster shapes and residual fluctuations in the gas density that weigh more in the PCA once the largest inhomogeneities caused by substructures have been removed. Contrary to intuition, the asymmetries introduced by these larger inhomogeneities lead to a more stable PCA near the virial radius by defining preferred directions of elongation. These, however, are nonetheless strongly misaligned from the median cluster orientation (on average by $20^{\circ}$ to $30^{\circ}$) and might not reflect the actual bulk orientation. }
}\\
	
	\item \hl{Substructure-cleaned data sets:\\Comparing \ch{the outcome for different values of} $X$, we note interesting differences between 2S, 3S and 4S. While the qualitative trends in the radial evolution are mostly the same, the magnitudes of $\epsilon, A$ and $\Delta \phi$ differ for each choice of $X$. This can be seen most prominently for the ellipticity, where the 2S curve lies significantly below the 3S result, e.g. the median deviates by \ch{typically} $8 \%$ from 3S (and $14\%$ from 4S) 
	For the aspheroidicity and the relatvie orientation, differences are smaller \ch{(around $4\%$ difference between the median values between 2S and 3S, and $10\%$ between 2S and 4S)}. 
	\ch{Thus, especially the} $X = 2$ criterion leads to more spherical cluster shapes by removing elongation in the direction of the major axis while less affecting the two minor axes. The low differences between 3S and 4S in combination with the large difference between 2S and all the other data sets at large radii suggest that $X = 2$ is likely too strict, erasing parts of the bulk component both at small and large radii, which is undesired and must be avoided.}
\end{itemize}
	\hl{\ch{We note a low deviation between all data sets at large radii ($r/r_{200} \gtrsim 0.8$) except for the large deviation of 2S from the other curves. The low differences even between 4S and the raw data set suggest that $X \gtrsim 3$ may not be efficient for removing all substructures at these radii. Indeed, when inspecting the isocontours of the substructure-cleaned gas density datasets, we note that a weak impact of inhomogeneities is still present and perturbs the shape of the gas density distribution.
	}} 

\subsubsection{Cluster shapes from the gravitational potential} \label{sec:relaxed_potential_ana}

\begin{figure*}[h!]
\begin{subfigure}[b]{0.328\textwidth}
  \centering
  \includegraphics[width=1.01\columnwidth,angle=0,clip]{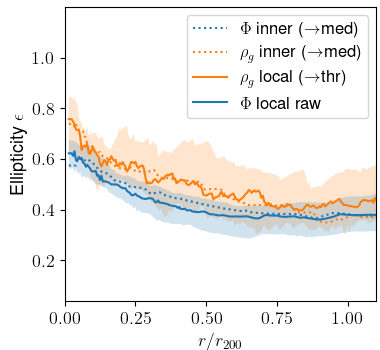}
   \caption{}
   \label{fig:relaxed_comp_1}
\end{subfigure}
\begin{subfigure}[b]{0.328\textwidth}
  \centering
  \includegraphics[width=1.01\columnwidth,angle=0,clip]{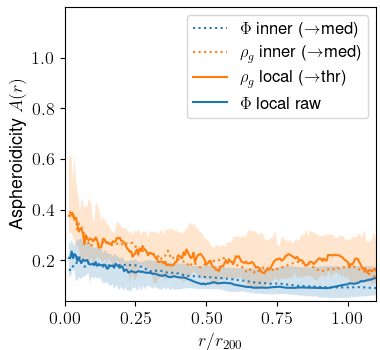}
  \caption{}
 \label{fig:relaxed_comp_2}
\end{subfigure}
\hspace{0.5pt}
\begin{subfigure}[b]{0.328\textwidth}
  \centering
  \includegraphics[width=1.01\columnwidth,angle=0,clip]{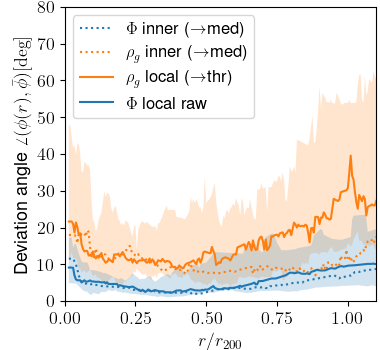}
  \caption{}
   \label{fig:relaxed_comp_3}
\end{subfigure}
\caption{\chn{Gas vs. potential --} \chn{Morphological parameters $\epsilon(r)$, $A(r)$ and deviation in orientation $\Delta \phi$} from the gravitational potential for relaxed clusters, and comparison to the gas density. \chn{The semi-transparent areas enclose the quartiles of the cluster sample, while “med” and “thr” denote whether regions occupied by substructures are replaced by the median or threshold approach (“a”, “b” and “ii” defined in Sect.~\ref{sec:subremoval}). The label “raw” denotes the non-processed original dataset.}}
\label{fig:relaxed_gas_vs_pot}

\vspace{1em}

\begin{subfigure}[b]{0.326\textwidth}
  \centering
  \includegraphics[width=\columnwidth,angle=0]{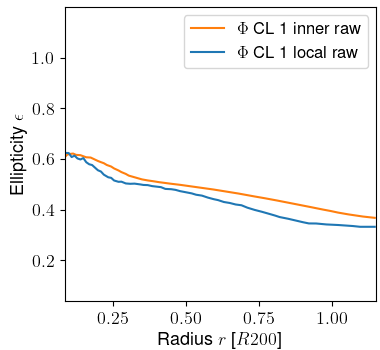}
  \caption{}
   \label{fig:relaxed_pot_indiv1}
\end{subfigure}
\hfill
\begin{subfigure}[b]{0.326\textwidth}
  \centering
  \includegraphics[width=\columnwidth,angle=0]{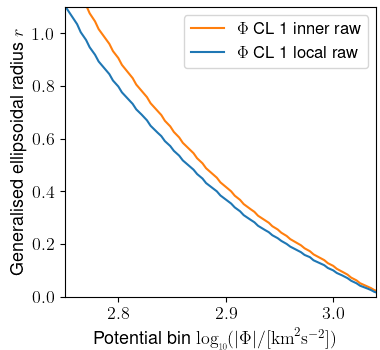}
\caption{}
   \label{fig:relaxed_pot_indiv2}
\end{subfigure}
\hfill
\hspace{1pt}
\begin{subfigure}[b]{0.326\textwidth}
  \centering
  \includegraphics[width=\columnwidth,angle=0]{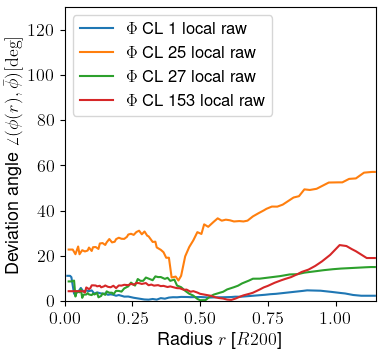}
  \caption{}
   \label{fig:relaxed_pot_indiv3}
\end{subfigure}
\label{fig:relaxed_pot_indiv}
\caption{Individual cluster result examples for the gravitational potential, showing the same random set of clusters as we used in Fig.~6. These figures are intended for a better illustration only, complementary to our statistical analysis using the full cluster sample. \chn{The label “raw” refers to the non-processed original potential data set.}}
\vspace{1em}
\begin{subfigure}[b]{0.328\textwidth}
  \centering
  \includegraphics[width=1.01\columnwidth,angle=0,clip]{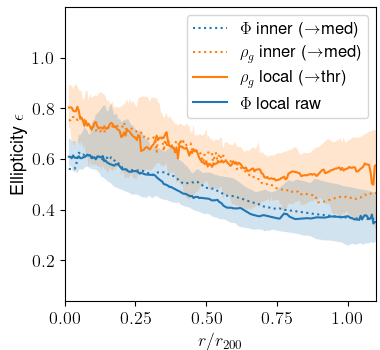}

   \caption{}
   \label{fig:unrelaxed_gas_domain1}

\end{subfigure}
\begin{subfigure}[b]{0.328\textwidth}
  \centering
  \includegraphics[width=1.01\columnwidth,angle=0,clip]{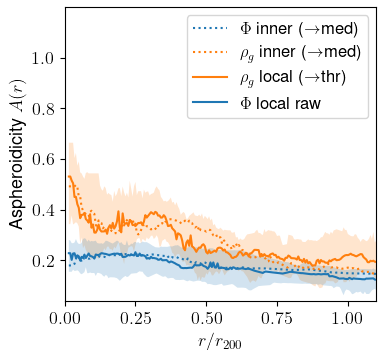} 
  \caption{}
 \label{fig:unrelaxed_gas_domain2}
\end{subfigure}
\hspace{0.5pt}
\begin{subfigure}[b]{0.328\textwidth}
  \centering
  \includegraphics[width=1.01\columnwidth,angle=0,clip]{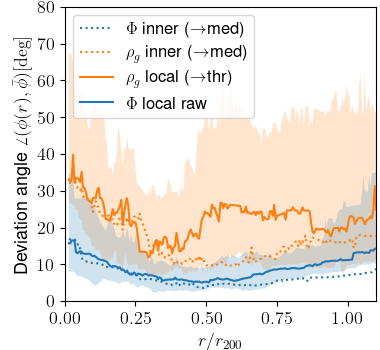}
  \caption{}
   \label{fig:unrelaxed_gas_domain3}

\end{subfigure}
\caption{Morphological parameters and deviation in spatial orientation for non-relaxed clusters, comparing gas density and potential. \chn{The semi-transparent areas enclose the quartiles of the cluster sample, while “med” and “thr” denote whether regions occupied by substructures are replaced by the median or threshold approach (“a” and “b” defined in Sect.~\ref{sec:subremoval}). The label “raw” denotes the non-processed original dataset.}} 
\label{fig:unrelaxed_gas_domain}
\end{figure*}

\hl{For the gravitational potential, we distinguish two different data sets: The \ch{non-processed (“raw”)} simulation output and a \ch{pre-processed data set} in which regions occupied by gas substructures (identified with $X = 3.5$) are substituted by the median potential in shells in formal analogy to the procedure for the gas density.
With the raw version, our aim is to compare whether the true, non-processed potential, including \ch{all its intrinsic asymmetries and} contributions from subclumps, still allows for a less ambiguous shape determination than the substructure-removed gas density. The second dataset, produced by substituting the median potential in substructure regions is included to provide \ch{a slightly different} comparison that aims \ch{to compare the} data sets pre-processed in the most similar way.
}
\hl{The results are shown for the local PCA method in Fig.~\ref{fig:relaxed_gas_vs_pot}, \ch{next to the results for the gas density, which are shown for a direct comparison.
We summarize and compare the results in the following:}}

\begin{itemize}
        \item \hl{ Ellipticity:\\
        The gravitational potential $\Phi$ (see Fig.~\ref{fig:relaxed_comp_1}) starts at a median ellipticity of $\epsilon_{\Phi} \sim 0.6$ in the core, which decreases down to a near-plateau at $\epsilon_{\Phi} \sim 0.4$ towards $r/r_{200} \gtrsim 0.5$. Thus, its behavior is qualitatively similar to the gas density, except that $\epsilon_{\Phi}$ is consistently smoother, slightly rounder than the gas (with a median ellipticity ca. $15\%$ to $20\%$ lower), and has much narrower inter-quartile range (by factors $\gtrsim$ 2).
        Thus, at fixed radius, the spread in $\epsilon_{\Phi}$ over the cluster population is considerably smaller than for the gas density\footnote{\ch{Note that when comparing individual clusters, the difference in fluctuation behavior becomes even larger because the sample averaging itself smoothes out many of the fluctuations in $\epsilon_{\rho_g}$, which are contributing to the large inter-quartile distances. For illustration, see the example in Fig.~\ref{fig:relaxed_pot_indiv1} in comparison to Fig.~\ref{fig:relaxed_gas_indiv1}}.}.
        \ch{The result is little sensitive to the PCA domain and holds even when comparing the local result for the potential with the substructure-removed inner domain result for the gas density. }
	Moreover, we note that $\epsilon_{\Phi}$ is nearly insensitive to whether we mask regions containing substructures (and fill them with the median method (ii).) Thus, the shape of the potential along its major axis is highly stable against local fluctuations in the data set. Another interesting difference between $\epsilon_{\Phi}$ and $\epsilon_{\rho_g}$ is that the potential curve remains flat around $r_{200}$ while the gas density ellipticity shows a slight increase due to growing asymmetries in the gas density distribution towards $r_{200}$. } \\

        \item \hl{ Aspheroidicity:\\
        From Fig.~\ref{fig:relaxed_comp_2} we can see that the gravitational potential is considerably closer to spheroidal symmetry than the gas density -- with the gas density showing on average twice as high values for $A$ than the potential. While for the gas density the median aspheroidicity starts from a central value slightly above $0.4$ and mostly remains fluctuating around $0.2$ to $0.25$ with higher radius, regardless of the method, the median aspheroidicity for both potential data sets starts slightly above $0.2$ and decreases very smoothly below $0.1$ until $r \sim 0.9 r_{200}$.
        In comparison to the gas density, $A(r)$ changes much more slowly with radius and with less fluctuations, and again the \ch{inter-quartile range is much narrower.} The upper quartile \ch{of $A_{\Phi}$ mostly} remains \ch{ even below } the median $\epsilon_{\rho_g}$ \ch{or just touches it}.
        Differences between the median-substituted and the \ch{raw} potential data set are \ch{again typically} very small, in contrast to the huge difference we \ch{noted between the 3S (or 4S) and raw gas density data sets} in Fig.~\ref{fig:relaxed_gas_domain6}.}\\

        \item \hl{ Relative orientation:\\
        The spatial orientation of the major axis computed from the potential \ch{shows considerably less fluctuations and deviation from the median orientation} than for the gas density data set. \ch{The angular difference $\Delta \phi$ to the median cluster major axis mostly remains below $10^{\circ}$ (median difference: $5^{\circ}$), while for the gas density it varies between $10^{\circ}$ and $30^{\circ}$ (median difference: $14^{\circ}$). The upper quartile for the potential varies between values of $10^{\circ}$ and $20^{\circ}$ (gas density: $20^{\circ}$ to $60^{\circ}$).
        The upper quartile is slightly lower as well, staying mostly below $50^{\circ}$ for the potential while the relative deviation from the median is typically around $40 \%$ higher for the gas density.
For the inner gas density data set, the deviation angle is mostly similar to that of the \chn{local method} -- only towards large radii, $r \gtrsim 0.7 r_{200}$ the deviation angle for the inner gas density data set is similar to that of the potential data sets due to the slow response of the so-defined mass tensor to local shape changes. Still, the corresponding value for the inner potential data set is considerably lower.
Thus, comparing only the inner, median-substituted data sets for gas and potential, the potential shows a more stable deviation angle at a smaller value than for the gas density.}}
\end{itemize}

\subsection{Cluster shapes of non-relaxed clusters} \label{sec:results_unrelaxed}

\hl{We now turn to the 53 non-relaxed clusters of the \textsc{Omega500} simulation to study their shapes and investigate the influence of the cluster's dynamical state by comparison with the relaxed sample. 
The non-relaxed cluster sample consists of objects showing signs of \ch{recent} dynamical activity in the past, \ch{such as irregular X-ray isophotes}, as defined \ch{through the SPA criteria} in Sect~\ref{sec:omega500_sim}.
For the \ch{shape parameters that we derive for the three-dimensional} matter and potential distributions, this may imply asymmetries, multiple mass peaks and \ch{higher} deviations between gas and potential isosurfaces. 
\ch{Due to the larger irregularity, the prevalence of substructures matching the $X$-$\sigma$ criterion is higher as well, such that the uncertainties and ambiguities related to substructures that we identified in the previous section might become even more important. In this section we fix the substructure criterion to the default choice of $X = 3.5$ to focus on the impact of the cluster dynamical state by comparing the results for the relaxed and non-relaxed samples. }
}

\begin{itemize}
	\item \hl{Ellipticity: \ch{The ellipticity profile, presented in} Fig.~\ref{fig:unrelaxed_gas_domain1}, \ch{shows that the gas density} ellipticity $\epsilon_{\rho_g}$ \ch{is consistently higher than for the relaxed sample for the gas density -- by around $15\%$. The inter-quartile distance \chn{also increases}. For the potential, the median ellipticity $\epsilon_{\Phi}$ is very close to its relaxed values, with the difference that it decreases more slowly with radius than for the relaxed sample, until it reaches the same plateau at $r/r_{200} \gtrsim  0.75$ as before. Interestingly, however, the inter-quartile distances are comparably large for the potential, with a pronounced peak in the upper quartile at $ r/r_{200} \sim 0.15$, touching the upper quartile of the gas density ellipticity. }
\ch{We note that in this region, the impact of the presence of multiple cluster cores appears to be largest, giving rise to a bi- or multimodality in the distributions of gas and dark matter -- while causing equipotential surfaces to become at least more elliptical (\chn{for two examples}, see Fig.~\ref{fig:snapshots_unrelaxed})} 
\ch{raises} the upper quartile of both gas density and potential ellipticities considerably. 
}\\

	\item \hl{Aspheroidicity: For the aspheroidicity, shown in Fig.~\ref{fig:unrelaxed_gas_domain2}, one can make an interesting observation: the median value for the potential increases only little \ch{compared to the relaxed sample}, staying below $25$ percent, whereas for the gas density, $A_{\rho_g}$ increases to values between $30\%$ and $55\%$ -- with broad quartiles and strong fluctuations while the potential is comparably smooth and nearly constant in $A_{\Phi}$ throughout large intervals in radii.
}\\

	\item \hl{ Spatial orientation: The change in spatial orientation of the major axis is shown in Fig.~\ref{fig:unrelaxed_gas_domain3}. Here, we can see a rather drastic increase and strong fluctuations in the results for the gas density, \ch{fluctuating between $30^{\circ}$ and $40^{\circ}$ in the median, with upper quartiles extending beyond $60^{\circ}$ over a large range in radii.} The potential, in contrast, still remains comparably stable in its spatial orientation, \ch{with median values between \ch{ $5^{\circ}$ and $18^{\circ}$ } and upper quartiles up to ca. $30^{\circ}$}.
	In contrast to the relaxed case, however, the potential shows a slightly higher level of fluctuations too.}

\end{itemize}

\section {Discussion}\label{sec:disc}

\hl{In the following we discuss the outcomes of our morphological analyses, their implications and their uncertainties. We comment both on physical implications and explanations of our results as well as discussing methodical aspects that lead to differences or ambiguities in the outcome.
We start by dicussing common trends that were visible in all data sets and then focus on the differences between single methods.
The latter we discuss in the context of removal of substructures, at the impact of our choice of PCA method and on the influence of the clusters' dynamical state, which reflects in the difference between the relaxed and non-relaxed populations.}

\subsection{Physical context and implications of cluster shapes}

\ch{
We found in this study that relaxed clusters have a median ellipticity of $\langle \epsilon_{\rho_g} \rangle \sim 0.5 \pm 0.1$ (for the gas density shells) and $\langle \epsilon_{\Phi} \rangle \sim 0.38 \pm 0.06$ (for the potential), when averaged within the interval $r \in [0, r_{200}] $. The deviation from spheroidal symmetry was quantified to be $\langle A_{\rho_g} \rangle \sim 0.21 \pm 0.08$ and $\langle A_{\Phi} \rangle \sim 0.11 \pm 0.06$. Thus, the gas density deviates by on average two times more from spheroidicity than the potential.
These values are consistent with the perturbative triaxial modelling of halo potentials by \cite{lee2003}, who found an analytic relation between gas density and mass axis ratios that allows to arrive at gas ellipticities of order $0.2$ to $0.5$; 
and at small values of the aspheroidicity below $0.2$. 
Our results for the potential shape parameters are in good agreement with those obtained from $N$-body simulated halos by \citep{hayashi2007}, who suggest $\epsilon_{\Phi} \sim 0.48 \pm 0.06$ near the halo center and less at higher radii; and $A_{\Phi} \sim 0.13 \pm 0.02$ remaining nearly constant over a large range in radius.
An earlier systematic study of 16 clusters simulated with the ART code by \cite{lau11b} show very similar values for both quantities in the non-radiative runs: $\epsilon_{\rho_g} \sim 0.4 ... 0.6$, $\epsilon_{\Phi} \sim 0.5 ... 0.6$, $A_{\rho_g} \sim 0.2 ... 0.4$ and $A_{\Phi} \sim 0.2 ... 0.3$.
Surprisingly, in their study, the potential partially has a higher ellipticity than the gas density and $\epsilon_{\Phi} $ increases again after $r/r_{200} \sim 0.3$ -- although the uncertainties are significant. 
}
\chn{Finally, our gas ellipticity measurements are consistent with those obtained earlier for a sample of 80 \textsc{Omega500} clusters by \citep{chen19}, who focused on the gas ellipticities and their correlation with mass accretion history, finding values $\langle \epsilon_{\rho_g} \rangle$ peaking around $0.6$, with scatter between $0.3$ and $0.8$. }

\hl{
One of the main trends we could identify in the morphological evolution of all data sets was that of an overall decrease in ellipticity and aspheroidicity of both gas and potential with radius.
Our study suggests that clusters become rounder and also more spheroidal with increasing radius, \ch{in agreement with earlier theoretical studies \cite[e.g.][]{hayashi2007,lau11b}}. Although \ch{in all comparisons, we found} the gravitational potential to be more symmetric \ch{than the gas density at most radii}, the qualitative behavior of ellipticity, aspheroidicity and major axis deviation is mostly very similar \ch{between} gas and potential. Quantitatively, differences can be noted in the (consistently) larger fluctuations and larger asymmetries present in the gas density distribution. These differences are notable for \ch{all} substructure-cleaned data sets and they become much larger when not removing substructures in the first place.}

\hl{The trend of isocontours getting more spherical with growing radii is expected for a gravitational potential generated by an ellipsoidal mass distribution. From potential theory it is well known that the exterior equipotential surfaces of thin homoeoids are confocal spheroids \citep{bt08} and that the monopole contribution dominates at large distances from a centrally concentrated mass distribution\footnote{This is valid at least as long as the effects of substructures and -- at very large radii -- mass contributions from filaments can be assumed to be of negligible order, as is the case for our data sets.}. }
\chn{In addition, earlier works have suggested that the shapes of cluster mass distributions themselves become slightly rounder towards the outskirts in numerical N-body simulations \citep[e.g.][and refs therein]{allgood06,veraciro11}.}
\hl{For the intracluster gas residing in a potential well generated by an \chn{arbitrary} mass distribution, we expect \ch{the gas density distribution to follow the potential closely in} regions that are in \ch{approximate} hydrostatic equilibrium.
The hydrostatic equation \ch{implies the gradients of gas density and potential} to be parallel,
$\nabla \rho_{\mathrm{gas}} \times \nabla \Phi = 0$, such that equipotential surfaces are common isosurfaces of density, pressure and X-ray emissitivity \citep[e.g.][]{buote94}. 
\chn{Thus, we expect the gas density to follow similar overall trends in orientation and axis ratios as the potential.}}

\hl{\ch{Nonetheless, the parameters} characterising potential \ch{shapes} are \ch{neither found nor expected to be} in full agreement with \ch{those of} the gas density (\ch{see} e.g. Figs.~\ref{fig:relaxed_gas_vs_pot} and ~\ref{fig:unrelaxed_gas_domain}). Even in ideal hydrostatic equilibrium, the gravitational potential $\Phi$ is expected to be distributed more smoothly than the gas density $\rho_g$. This can be seen explicitly for an ideal or polytropic gas with an effective adiabatic index $\gamma$, for which the hydrostatic equation implies the following analytical relation \citep{konrad13}: 
\begin{align}
	\Phi \propto \rho_g^{\gamma-1}	
\end{align}
For realistic adiabatic indices, $1.1 \lesssim \gamma \lesssim 1.2$ \citep{finoguenov01}, the \ch{small} exponent \ch{causes} potential fluctuations \ch{to be} suppressed \ch{relative} to density fluctuations by factors of 5 to 10:
\begin{align}
	\frac{\delta \Phi}{\Phi} \propto (\gamma - 1 ) \frac{\delta \rho_g}{\rho_g}.
\end{align}
This leads to a smoothing \ch{that can be compared to a low-pass filter suppressing} small-scale components in the potential distribution, which \ch{therefore weigh less in} the mass tensor.
}

\hl{The gas density, \ch{however, contains significant} fluctuations even for relaxed clusters and after excluding substructures \ch{with the advanced methods such as the $X$-$\sigma$ criterion}. 
\ch{Some of these fluctuations are consequences of small deviations from the hydrostatic assumption. Due to the large typical dynamical and thermal timescales of galaxy clusters in comparison with the age of the Universe, hydrostatic equilibrium in galaxy clusters is only approximate.}
With a free fall timescale similar to that of sound crossing, the intracluster medium \ch{can be transsonic over large regions}, causing slight shocks \ch{across the cluster} that lead to discontinuities in the density.
\ch{The presence of such bulk fluctuations is expected to depend only little on the input gas physics \citep{zhuravleva13, roncarelli13}.}  
\ch{At the same time, further} deviations from hydrostatic equilibrium are \ch{expected \chn{due to} dynamical acitivity (i.e. accretion and \chn{merging activity} of substructures) at the outskirts, $r \gtrsim r_{500}$.}
In the cluster core regions, moreover, the effects of radiative cooling and baryonic feedback -- not included in this simulation -- may affect the shapes of clusters, with radiative cooling in the centers expected to lead to a more spherical gravitational potential due to a more centrally condensed mass distribution \citep[e.g.][]{Dubinski94, Kazantzidis04}, while at the same time, AGN feedback may largely counteract this effect by preventing the gas from cooling \citep{bryan13, suto17, chua19}.}
\ch{Still, the details of these mechanisms need to be better explored and understood. When reconstructing cluster shapes in practice, we expect that deviations from the assumption of hydrostatic equilibrium may be a larger source of concern than the impact of input gas physics on the actual shapes within the inner cluster region.}

\begin{figure}
\begin{subfigure}{\columnwidth}
  \centering
  \includegraphics[width=\columnwidth,angle=0]{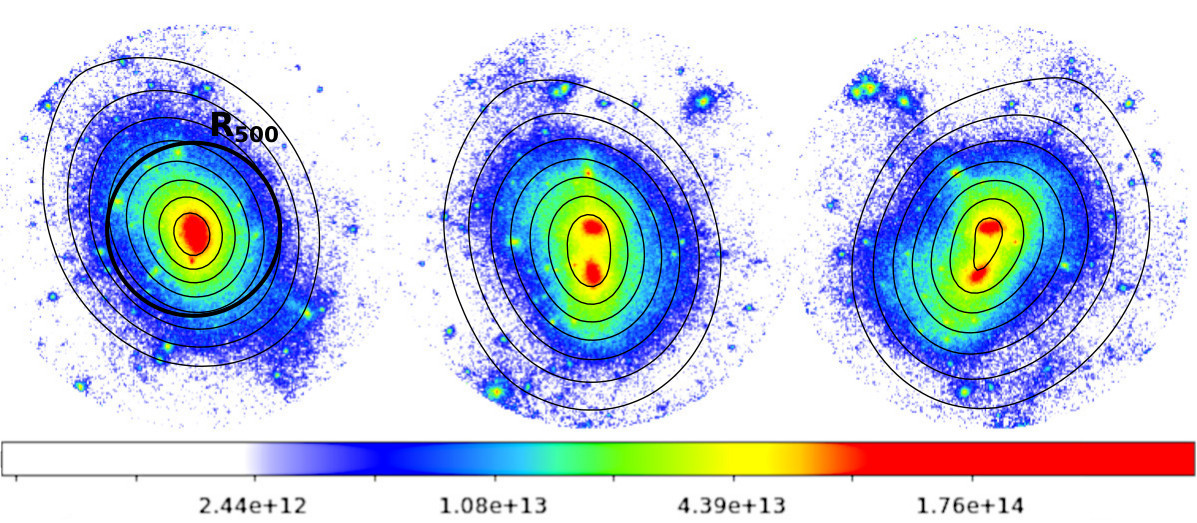}
	\caption{Cluster CL10}
 \label{fig:snapshotcl10}
\end{subfigure}

\begin{subfigure}{\columnwidth}
  \centering
  \includegraphics[width=\columnwidth,angle=0]{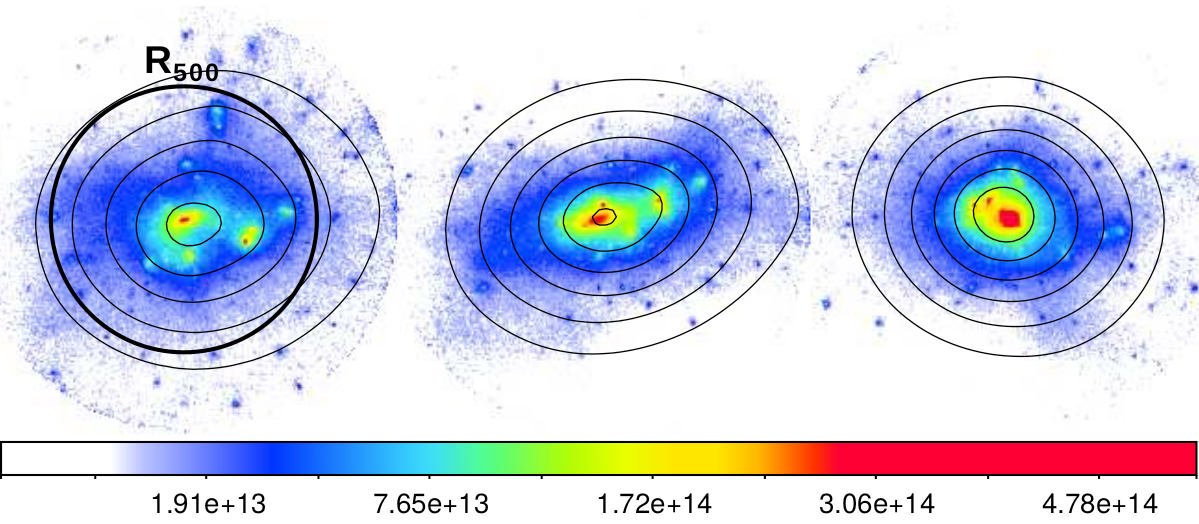}
  \caption{Cluster CL77}
 \label{fig:snapshotcl77}
\end{subfigure}
  \caption{Potential isocontours overlaid on 2D projections of the dark matter distribution for two non-relaxed clusters showing bi- or multimodality in their mass distribution.
The three panels in each case show projections of the simulated data cube for the three spatial orientations $x,y,z$ of the cube.} \label{fig:snapshots_unrelaxed}
\end{figure}
v

\subsection{Impact of dynamical state}

\hl{For the non-relaxed cluster sample, we \ch{noted that clusters have both more ellipticial and more triaxial shapes especially when looking at the gas density distribution. In addition, we found slightly} larger quantitative differences between gas and potential (see e.g. Fig.~\ref{fig:unrelaxed_gas_domain}) -- as was to be expected given that the clusters \ch{are dynamically more active as indicated by their irregular X-ray morphologies. In such cases, large parts of the gas may not yet have settled into approximate hydrostatic equilibrium, leading to more pronounced clumpiness, shocks and asymmetries.}
\ch{The presence of two or multiple, merging cores can, in addition, lead to more elliptical potential isosurfaces, for which two examples are shown in Fig.~\ref{fig:snapshots_unrelaxed}.}}
The increased ratio of ellipticity to aspheroidicity \ch{of the potential} suggests that a large number of mergers have their mass centers separated along (or near) the major axis. This is consistent with the predictions of gravitational collapse of dark matter driven by self-gravity for an initial Gaussian random field, which implies that cluster halos are elongated along the direction of their most recent major mergers due to the anisotropic accretion and merging along filamentary structures \citep[e.g.][]{Limousin13}.
\hl{The \ch{implication for the ellipticity of the potential became apparent from Fig.~\ref{fig:unrelaxed_gas_domain1}, where we saw increased} upper quartiles for the potential, \ch{although the latter is still very smooth and deviates significantly less from spheroidal symmetry than those of the gas density.} 
}



\subsection{Impact of methods and data sets} \label{methods_and_datasets}

\begin{figure}
  \centering
  \includegraphics[width=0.9\columnwidth,angle=0]{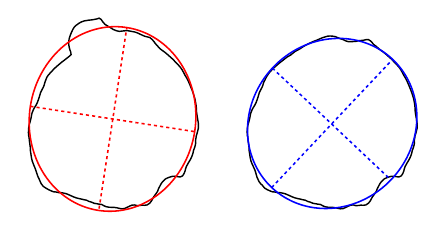}
  \caption{Illustration of the effect of fluctuations on the principal axis of a density isocontour. This example shows isocontours taken from the 3S (left) and 2S (right) datasets of a relaxed cluster when computing a slice of the cube.}
   \label{fig:2Dellips_fluct}
\end{figure}

\hl{In the previous section we discussed the impact of density inhomogeneities on the gas density. The presence of inhomogeneities and asymmetry, whether caused by non-virialised subhalos or residual shocks in the bulk gas component, complicates the determination of cluster shapes by introducing ambiguities and uncertainties, which are affecting the gravitational potential \ch{much less}. Apart from these difficulties \ch{when using the gas density}, the three-dimensional morphology of clusters can be defined in several different ways that may be ambiguous with respect to the choice of defining isodensity bins, summation volumes and radial profiles.}


\begin{figure}
\begin{subfigure}[b]{0.99\columnwidth}
  \centering
  \includegraphics[width=0.7\columnwidth]{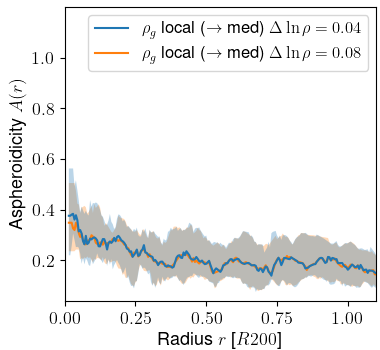}
  \caption{}
   \label{fig:bin_width_1}
\end{subfigure}
\hfill
\begin{subfigure}[b]{0.99\columnwidth}
  \centering
  \includegraphics[width=0.7\columnwidth]{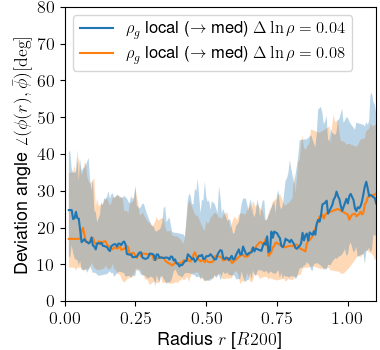}
  \caption{}
   \label{fig:bin_width_2}
\end{subfigure}
\caption{ Impact of the bin width on shape parameters quantifying asymmetry and orientation. The label “med” denotes that regions occupied by substructures are replaced by the median (“a” defined in Sect.~\ref{sec:subremoval}.}
\label{fig:fig_bin_width}
\end{figure}


\begin{itemize}
	\item Substructures and their removal:\\
	Substructures are density inhomogeneities that are not virialised with the cluster \hl{and therefore need to be identified and removed prior to} estimating the hydrostatic mass (\hl{or} potential) based on the gas density distribution \citep[e.g.][see also our companion paper \citealp{tchernin20}]{nagai11, zhuravleva13, roncarelli13, vazza13}.
Additional analysis based on the abundance/temperature of these clumps could be used to distinguish the substructures from the cluster body. However, such information is not always available as it requires high resolution measurements. Nowadays, there exist \hl{several} different ways to \ch{identify or suppress} substructures\ch{. For instance, \cite{Dubinski91} have applied an $r^{-2}$ weighting in the mass tensor to suppress the impact of dark matter subclumps, which are most prevalent in the dynamically younger cluster outskirts -- see however \cite{zemp11}, who suggested that this introduces systematic biases. Another definition of substructures is based on the (residual) gas clumpiness factor $\rho^2/\rho$ \citep[e.g.][and refs therein]{roncarelli13} or uses the more advanced $X$-$\sigma$ criterion (with $X = 3.5$, \citealp[e.g.][]{zhuravleva13, lau15}), which takes into account the lognormal shape of the bulk gas distribution.}
\ch{Applying such methods, however,} may alter the intrinsic shape of the cluster, which is the information we actually want to extract.

\hl{We therefore investigated} the systematic effect coming from such a treatment of data on the outcome of the \hl{shape reconstruction}.
\hl{For this} we compared \hl{\ch{several different generalisations of the $3.5 \sigma$-method, which we dubbed the $X$-$\sigma$ method}: Apart from the $X = 3.5$ criterion, we produced versions of the gas density data set for $X=2$, $X= 3$ and $X=4$ as well as for the raw, non-processed gas density distribution to highlight the effect of substructure-removal on the final results.
It should be noted that the choice for the cut-off parameter $X$ is fully arbitrary and needs to be chosen carefully to avoid erasing too large parts of the bulk component.}

\hl{This introduces two notable sources of ambiguity: (1) it is unclear which value should be chosen for $X$ as there is neither a sharp distinction between bulk and substructure X-ray signals nor a way to distinguish whether asymmetries in the density isolayers are caused by substructures or intrinsic to the bulk (triaxiality-substructure degeneracy); (2) it is unclear what to substitute for the regions occupied by substructures.
Even after cleaning substructures, we are left with residual fluctuations in the bulk component, e.g. due to small shocks present in the gas. These fluctuations are present on all scales and affect observable quantities like the X-ray emission, which depends on the square of the gas density. 
{In contrast to the decreasing ellipticity and aspheroidicity, we observed that the relative deviation in orientation of the major axis generally increases with radii. This is most likely the result of two \chn{combined effects:} First of all, substructures become important at large radii, affecting the derived shape parameters. Second, the more spherical the intrinsic cluster shape is, the less well the orientation of the three axes is constrained and the more sensitive it becomes to any fluctuations along the contours of the ellipsoid. This phenomenon, which is illustrated in Fig.~\ref{fig:2Dellips_fluct}, affects the stability of the PCA.}}

\hl{This is a problem that does not affect the gravitational potential, which we observed to be nearly insensitive to the contributions of sub-clumps -- thus enabling the use of the raw, non-processed potential data for the shape determination without any problems.
Due to its suppression of small-scale components, the potential distribution is intrinsically smoother and more symmetric than the gas density.}\\

	\item PCA domain:\\
	We applied the PCA method to \hl{simulated clusters for different definitions of the integration domain for the mass tensor} (\hl{`inner' vs `local'}, see Fig.~\ref{fig:methods3DPCA}).	
	We noticed that information derived \hl{from the local domain is \ch{more} sensitive to fluctuations in the gas density, \ch{as it responds to any local changes in the shape of isocontours}.}
\ch{The inner domain, in contrast, is more robust, as expected due to the sum over larger regions enclosed by the isocontours. Its disadvantage, however, is that exactly this property makes it respond slowly to the actual radial evolution of cluster shapes, similar to the observations made by \citep{zemp11} for different choices of defining a halo shape tensor. Therefore it does not capture some features that may be intrinsic and of interest. Moreover, it develops a non-trivial dependence on the inner regions such that shape features are difficult to associate to a specific radius -- and the resulting shape measurements depend on the slope of the gas density profile. }
\ch{In our study, we saw that the differences between the two methods become notable at larger radii at least when looking at the gas density.}
The absolute level of fluctuations and uncertainty in ellipticity, aspheroidicity and orientation angle is \ch{considerably larger for the local method. \chn{In contrast,}  the difference between the two domains is found to be remarkably small for the gravitational potential. Thus, the potential-based PCA is much more robust against the domain definition than the gas-based shape reconstruction.}\\

\item \hl{Uncertainties due to bin and radius definition:\\ The radial evolution of shape quantities depends only weakly on the choice of bin width. \ch{We explored the impact of the bin width on our results by comparing different choices for the logarithmic bin width parameter. An example is shown for illustration} in Fig.~\ref{fig:fig_bin_width}, where we can see that even a twice as large bin width causes fluctuations of the quartiles in the orientation angle that are still of similar size or slightly below that of the original value.
In contrast, the aspheroidicity does not change significantly except for the innermost radii. \ch{Our tests using different bin width parameters suggest that} a large part of \ch{the fluctuations we found in the parameters characterising the shape and orientation of the gas density distribution} are not \ch{caused by a too narrow} bin width.}

\hl{The noisy fluctuations present in the radial evolution of the major axis orientation of the gas density \ch{lead to} the question whether these fluctuations are entirely intrinsic or whether a part of them \ch{might be} caused by the definition of the generalised ellipsoidal radius $r$. The latter is sensitive to fluctuations in the axis ratios, which in principle can lead to small random displacements of the plotted quantities along the $r$-axis. In order to quantify how stable the ellipsoidal radius is with respect to axis ratio fluctuations both with radius and between different clusters, we investigate the \ch{behavior of the} radius $r$ as a function of the density bin \ch{for the individual cluster data sets}. \ch{We find that the resulting curves $r(\log_{10} \rho)$ are in all cases rather smooth. An example is shown for illustration in Fig.~\ref{fig:relaxed_gas_indiv2} and Fig.~\ref{fig:relaxed_pot_indiv2}.
This suggests that the effect intrinsic fluctuations of iscontours have on the generalised ellipsoidal radius $r$ are small, not significant enough to cause artificial fluctuations in $\epsilon(r)$, $A(r)$ and $\angle(\phi(r), \bar \phi)$ purely due to random shifts along the $r$-axis. It should be pointed out, however, that radial fluctuations in $\epsilon(r)$ and $A(r)$ are naturally connected to changes in the axis orientations -- e.g. small random fluctuations in the axis directions might be visible in the length ratios too. These, however, should to a large part average out when taking the sample median. 
Still, as Figs.~\ref{fig:relaxed_gas_domain3},~\ref{fig:relaxed_gas_indiv3} and~\ref{fig:unrelaxed_gas_domain3} show, the large fluctuations indicating instabilities in the PCA are potentially problematic for measuring the shape of the gas density distribution. The values for $\epsilon_{\rho_g}$ and $A_{\rho_g}$ may thus be affected by larger uncertainties than indicated by the quartiles only. This problem does not significantly affect the potential, whose orientation is highly stable for the case of most clusters.
}}
%
%


\end{itemize}




\begin{figure}
  \centering
  \includegraphics[width=\columnwidth,angle=0]{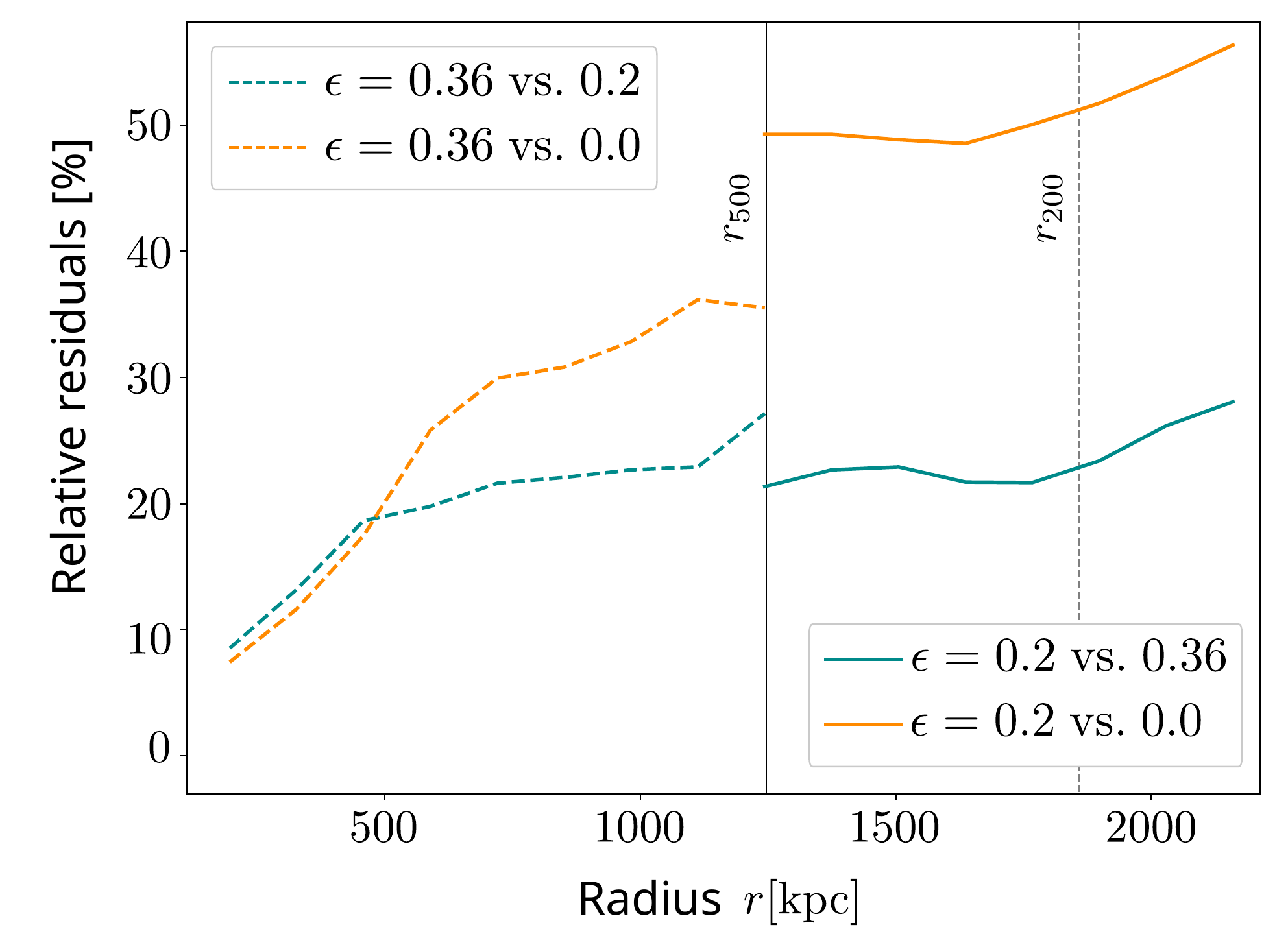}
  \caption{Relative residuals in X-ray emissivity profile \chn{for the example relaxed simulated cluster CL135 studied by \citet[][]{tchernin20}}, for the region \chn{[0.5, 1]} $r_{500}$ ($\epsilon \approx 0.36$), compared to the optimal ellipticity in the region \chn{[$r_{500}$, $r_{200}$ } ($\epsilon \approx 0.2$). \chn{These optimal ellipticity values are obtained using the inner domain.} For comparison, the relative residuals with respect to the spherical symmetry assumption are also shown. Here $x$ versus $y$ means: $\left|x-y\right|/x$, with $x$ and $y$, the triaxial profiles derived for a given ellipticity. The vertical lines indicate $r_{500}$ and $r_{200}$, respectively.}
   \label{fig:XRprofile}
\end{figure}

\subsection{Importance of accurate shape assumptions}

\hl{In the context of our comparisons between the symmetries of gas and gravitational potential we} note that it is crucial to model the 3D shape of a cluster as correctly as possible \hl{when working} with quantities like the gas density or the X-ray emissivity. This\hl{, however, is} not trivial, even for the most relaxed clusters, as our results show.
\hl{However, we find that even } small variations in ellipticity can lead to significant errors in the resulting triaxial profile. \hl{To illustrate this point for an example, we extracted a triaxial profile of the}  X-ray emissivity for the  cluster CL135 \chn{whose mass and potential profiles we studied in detail in \citep[][]{tchernin20} }, \hl{assuming} three different values for the ellipticity:
the optimal ellipticity \chn{obtained for CL135 from our PCA in the range $[0.5; 1] r_{500}$: $\epsilon=0.36$; the optimal ellipticity for radii larger than $r_{500}$: $\epsilon=0.2$; and the spherical assumption: $\epsilon=0$ (see Fig.~\ref{fig:XRprofile})}.
As we can see, a small variation of the ellipticity (between \chn{$0.2$ and $0.36$}) leads to \hl{an average} deviation of about 25\% over the entire cluster \hl{radius} range. \hl{Moreover,} assuming spherical symmetry for a cluster with an ellipticity as low as \chn{$0.36$ (in the range [0.5;1]  R$_{500}$) and $0.2$} (in the outskirts) can lead to underestimating the X-ray emissivity profile by up to 50\% in the region between $r_{500}$ and $r_{200}$, and by 30\% between \chn{$[0.5;1]  r_{500}$, which is part of the region accessible to X-ray measurements.
This effect can have non-negligible consequences on the study performed with these profiles (e.g.} in HE  mass estimates).

\hl{Similar} impacts due to over-simplified cluster shapes are expected to be numerous, even though they are difficult to quantify, given that the actual shapes of clusters are unknown. For instance,  using a technique like the median approach \citep{eckert15} to remove substructures in spherical shells in a spheroidal cluster can lead to underestimating the resulting profile by up to 15\% \citep[see Fig.~A.2a and A.8 in ][]{tchernin20}.

\section {\chn{Summary \& conclusions}}\label{sec:conc}

\hl{The three dimensional structure of galaxy clusters contains important information on the initial density field, the halo assembly, \chn{cosmological parameters and the nature of dark matter and gravity. As such, it will play a crucial role for upcoming cosmological studies.}
%
\chn{Information on the gas density distribution obtained from X-ray/SZ observations  has been extensively used to  draw cosmological constraints. However, we saw that deriving the shape of a galaxy cluster is not trivial even for a relaxed cluster, from which we have all the simulated data at hand.}}
Based on the gas density distribution \chn{of relaxed simulated galaxy clusters}, we emphasize that deriving the shape of a galaxy cluster is not trivial even from the full 3D simulation. This is due to the fact that cluster shapes evolve with time and therefore, that their morphology \hl{can vary} with the distance to the center. The presence of substructures adds to the complexity of the problem, as \hl{their exclusion introduces new uncertainties and ambiguities, e.g. using} a too strict criterion may alter the \hl{reconstructed} shape of a cluster.

In the present study, we analysed \hl{the shapes of 85 relaxed and non-relaxed clusters of galaxies} from the {\textsc{Omega500}} simulation.
\hl{We highlighted the difficulty of removing substructures from the matter density distribution and showed how this affects the reconstructed shape of the cluster estimated from the mass tensor.
We also illustrated the consequences a wrong assumption on the cluster shape can have on the resulting extracted X-ray emissivity profiles derived for a low ellipticity relaxed cluster.}
\hl{As a result, we} arrive at two main conclusions \hl{that we summarise in the following two subsections.}

\subsection{\chn{Shortcomings of the gas density as a shape proxy}}

Our results show that the distribution of \hl{the gas} density in clusters of galax\hl{ies} is a weak proxy for the cluster shape.
\hl{This is because the gas density contains large inhomogeneities due to substructures, which need to be excluded and are partially degenerate with the triaxiality of the cluster. In addition, different methods exist} both for identifying and removing substructures. \hl{Even when substructures are excluded, fluctuations in the bulk gas density complicate the morphological analysis.}
   We tested here: (1) how the result of the PCA is affected by the method used to remove substructures 
   (2) the \hl{sensitivity of the PCA to perturbations in the resulting isodensity shells quantified by the radial fluctuation behavior;} 
   (3) how the \hl{PCA domain (`inner' vs `local' domain) influences} the outcome of \hl{the cluster shape determination;} \chn{and (4) by how much the cluster shape parameters and orientation varies between different clusters}.

Clusters are often assumed to have a simple geometric shape \hl{(spherical or spheroidal)} for extracting various \chn{types of} information from the 2D observations and \hl{linking} these to theoretical models.
The shape of the \hl{gas} density distribution is \hl{more} difficult to characterise \chn{by using} a simple model, \hl{deviating strongly from spherical symmetry $\langle \epsilon_{\rho_g} \rangle \sim 0.5 $ and significantly from spheroidicity $\langle A \rangle \sim 0.21$.}
\hl{Given the intrinsic asymmetries and inhomogeneities in the gas density even after removing substructures, the use of spheroidal models not only leads to a loss of information -- but in addition, the gas density distribution allows for a less stable PCA.}
\hl{Both the shape and the orientation of isodensity shells measured for the gas density distribution vary significantly with radius, analysis method, substructure-cleaning method, dynamical state and between different clusters.} 
   For instance, the PCA applied to data cleaned from substructures with the $3.5 \sigma$ criterion \hl{leads to large deviations both between different clusters and with respect to other choices of the substructure identification parameter $X$. In addition, shape quantities derived from the mass tensor are subject to significant fluctuations with radius, for example in the orientation angle.}
   Therefore, studies of cluster shapes based on the gas distribution in galaxy clusters are not robust.

    We also \hl{highlighted} that it is crucial to properly approximate the 3D shape of a cluster \hl{when} dealing with quantities like the gas density or the X-ray emissivity. \hl{Even} small variations in ellipticity can lead to significant errors in the resulting triaxial profile (see Fig.~\ref{fig:XRprofile}). 
    In addition, we observed in our companion paper \citep{tchernin20} that using a technique like the median approach \citep{eckert15} to remove substructures in spherical shells on a spheroidal cluster can bias low the resulting profiles. This implies that systematical errors arising from substructure removal should be taken into account in the extracted X-ray profiles. Such systematics will propagate into other quantities derived from these profiles, e.g., into HE mass estimates.\\

\subsection{Advantages of using the gravitational potential}

\hl{The gravitational potential is in multiple ways better suited as a \chn{shape and profile} reconstruction quantity \chn{than the gas density}.
Not only \chn{does it enable a simpler, less ambiguous definition compared to cluster masses \citep[given its nature as a local quantity directly related to observables, e.g.][]{angrick09,lau11,angrick15} and lower systematics} in \chn{profile} reconstructions on simulated data (e.g. $6\%$ bias instead of $13\%$ for the HE mass; and a scatter reduced by $\sim 35\%$, see \citealp{tchernin20}) -- but the results of our study suggest that the shape of the gravitational potential is better approximated by a simple geometry (e.g. spheroidal symmetry) without significant loss of information. We also find that, due to its smoothness\footnote{Which results from the suppression of small-scale modes compared to the density distribution \citep[see also section 2 in][]{tchernin20}.}, the potential allows for a more robust PCA, avoiding several problems and ambiguities in determining cluster shapes.}

\chn{We noted that even for dynamically relaxed clusters in approximate hydrostatic equilibrium, the characterisation of galaxy cluster shapes based on the gravitational potential $\Phi$ has significant advantages over the gas density, in particular:}

\begin{itemize}
	\item
\hl{Simple geometry, low scatter:}\\
\hl{Cluster gravitational potentials can be better approximated by a simple spheroidal model -- with a typical aspheroidicity $\langle A_{\Phi} \rangle \sim 0.11$ twice as low than the gas density, a slightly lower ellipticity $\langle \epsilon_{\Phi} \rangle \sim 0.38$, and with less scatter (in all quantities) among the relaxed\footnote{For the non-relaxed sample, see the item “dynamical state” below.} cluster population.}
Studies aiming at reconstructing the potential from observations to constrain theoretical predictions are therefore significantly less affected by simplifying assumptions about its shape \hl{than if using the gas density}.
   As 2D observations always need to be deprojected to be related to theoretical predictions, this \hl{allows for a much more straightforward cluster characterisation.}
   \hl{Still, all information about the underlying mass distributions (both gas and dark matter) and on common observables (X-ray, lensing, SZ, galaxy kinematics) are encoded in the gravitational potential \citep[see e.g.][]{konrad13,sarli14,stock2015,majer16}.\\}

\item \hl{Smoothness \& stability against fluctuations:}\\ 
\chn{Since the gravitational potential is a considerably smoother quantity than the matter density distribution, its isoshells are nearly unaffected by local perturbations of any kind in the mass distribution -- even though it includes the full\footnote{Or partial, when masking regions in which gas substructures where identified.} impact of non-virialised substructure contributions $\delta \Phi$. Especially the spatial orientation of potential isolayers remains highly stable with increasing ellipsoidal radius ($\Delta \phi_{\Phi} \sim 5^{\circ}$ to $10^{\circ}$ for relaxed clusters). This shows that the cluster orientation is relatively constant with radius; and this needs to be contrasted with the large variations we measured for the gas density ($\gtrsim 10$ to $30^{\circ}$ and much larger in individual cases).}
\hl{Fluctuations present in the mass density distribution are smoothed out when integrating Poisson's equation to derive the gravitational potential.}
\chn{Still, the gravitational potential encodes all physical information about the underlying mass distribution.}
	\hl{The \chn{extremely low impact of inhomogeneities on the potential-based PCA implies a more reliable physical characterisation of galaxy clusters, with less uncertainties in shape parameters. In particular,} no pre-processing step \chn{for substructure removal is needed, which require additional assumptions.}
	\chn{In this context, ambiguities regarding cut thresholds and filling methods are entirely avoided. Thus, the gravitational potential encodes the shape information of the underlying mass distribution in a smooth quantity, which (in projection) can be straightforwardly linked to many different observables.}}\\

	\item Dynamical state:\\
	 \hl{Even for non-relaxed clusters, the potential $\Phi$ can be approximated by a spheroid with a relative deviation $\lesssim 20\%$ from the actual shape.} This result is very encouraging. Combined with the fact that the potential can be relatively well reconstructed for clusters in any dynamical state  \citep[with a total bias of 6\%, as demonstrated in our companion paper][]{tchernin20}, 
this implies that by characterizing galaxy clusters by their gravitational potential in cosmological studies, we can include dynamically active clusters. This will help tighten the existing constraints on the cosmological models.
	Many cosmological studies rely on relaxed galaxy clusters only \citep[e.g.][]{vikhlinin09,mantz16,applegate16}, owing to the fact that \chn{these are more easy to compare} to theoretical predictions due to their expected spherical (or spheroidal) shapes. However, taking into account dynamically active clusters in such studies will increase the \chn{sample} statistics. \hl{Moreover, as it is generally difficult to assess whether an observed cluster is dynamically active or not, including these clusters will also significantly simplify the process of selecting the clusters.} 
This will be possible by replacing the mass of clusters by  their gravitational potential in such studies \citep[see also][]{lau11}, both in theoretical predictions \citep[using a "potential function" as for instance in][]{angrick15} and \hl{in observations, where the potential can be reoconstructed from direct observables like lensing, X-ray or millimeter observations \citep[as done in e.g.][]{tchernin15,tchernin18,merten16,huber19}}
%
The smooth, \chn{nearly spheroidal} shape of the potential (even for dynamically active clusters) will be an important advantage for this purpose.


%


\end{itemize}

\hl{Characterising galaxy clusters by their gravitational potential in cosmological studies will allow to improve constraints on cosmological parameters, the nature of dark matter, dark energy and gravity.}
\hl{The prospects for releasing the full power of potential-based cluster reconstructions for cosmological studies are highly promising given that large homogeneous and high-quality cluster datasets are or will become available across the electromagnetic spectrum in the near future.
X-ray observations e.g. from XMM-\emph{Newton} or \emph{Chandra} provide high-quality constraints for individual cluster reconstructions \citep[e.g.][]{merten11,jauzac16,hallman18,pandge19,jauzac19}. In combination with other probes, such as gravitational lensing from optical or SZ from millimeter observations e.g. from \emph{Planck}, SPT, ACT or IRAM, these observations can be used in multi-wavelength analyses to probe the physical conditions of clusters out to large radii \citep[e.g.][]{morandi12, xcop, ghirardini18, tchernin18, siegel18, huber19}.
The next decade of cluster cosmology will highly benefit from advanced large-scale surveys such as eROSITA \citep{merloni12}, Euclid \citep{laureijs11} or LSST \citep{lsst09}, which are going to yield order-of-magnitude increases in cluster samples, thus transforming our our ability to use galaxy clusters as cosmological probes dramatically.
It will be highly interesting to see what improvements and new insights can be achieved by using a potential based modelling approach. }
\acknowledgements{
	\ch{We thank the anonymous referee for valuable comments that helped to greatly improve this manuscript. 
	\chn{We} would like to thank Matteo Maturi, Sara Konrad and Jonas Große Sundrup for stimulating discussions and feedback.}
	This paper was supported in part by the Deutsche Forschungsgemeinschaft under BA 1369 / 28-1 and by the Swiss National Science Foundation under P2GEP2 159139. \ch{Sebastian Stapelberg and Matthias Bartelmann's work is funded by the Deutsche Forschungsgemeinschaft (DFG, German Research Foundation) under Germany's Excellence Strategy EXC 2181/1 - 390900948 (the Heidelberg STRUCTURES Excellence Cluster). }
	We thank Daisuke Nagai for kindly giving us the permission to use the data from the Omega500 simulation. The Omega500 simulations were performed and analyzed at the HPC cluster Omega at Yale University, supported by the facilities and staff of the Yale Center for Research Computing.}

\end{document}